\documentclass[twoside,twocolumn,9pt]{article}
\usepackage{extsizes}
\usepackage[super,sort&compress,comma]{natbib} 
\usepackage[version=3]{mhchem}
\usepackage[left=1.5cm, right=1.5cm, top=1.785cm, bottom=2.0cm]{geometry}
\usepackage{balance}
\usepackage{times,mathptmx}
\usepackage{sectsty}
\usepackage{graphicx} 
\usepackage{lastpage}
\usepackage[format=plain,justification=justified,singlelinecheck=false,font={stretch=1.125,small,sf},labelfont=bf,labelsep=space]{caption}
\usepackage{float}
\usepackage{fancyhdr}
\usepackage{fnpos}
\usepackage[english]{babel}
\addto{\captionsenglish}{%
  
}
\usepackage{array}
\usepackage{droidsans}
\usepackage{charter}
\usepackage[T1]{fontenc}
\usepackage[usenames,dvipsnames]{xcolor}
\usepackage{setspace}
\usepackage[compact]{titlesec}
\usepackage{hyperref}

\usepackage{epstopdf}

\def\eb{\begin{equation}}   
\def\ee{\end{equation}}     
\def\ea#1{\begin{eqnarray} #1 \end{eqnarray}}   

\def\ket#1{\bigl|#1\bigr>}
\def\bk{\hskip -3.0pt \bigm| \hskip -3.0pt}

\def\inprod#1#2{\bigl< #1 \bk #2 \bigr>}

\def\ep{\epsilon}

\def\of#1{\left(#1\right)}

\def\eq#1{Eq.~(\ref{#1})}
\def\eqs#1#2{Eqs.~(\ref{#1}) and (\ref{#2})}

\def\sof#1{\left[ {#1} \right]}

\newdimen\amstexex
\def\dotsadjustbox#1{\vbox to -1.4\amstexex{\kern-2\amstexex\hbox{#1}\vss}}

\def\ep{\varepsilon}

\definecolor{cream}{RGB}{222,217,201}

\begin{document}

\pagestyle{fancy}
\thispagestyle{plain}
\fancypagestyle{plain}{

}

\makeFNbottom
\makeatletter
\renewcommand\LARGE{\@setfontsize\LARGE{15pt}{17}}
\renewcommand\Large{\@setfontsize\Large{12pt}{14}}
\renewcommand\large{\@setfontsize\large{10pt}{12}}
\renewcommand\footnotesize{\@setfontsize\footnotesize{7pt}{10}}
\makeatother

\renewcommand{\thefootnote}{\fnsymbol{footnote}}
\renewcommand\footnoterule{\vspace*{1pt}%
\color{cream}\hrule width 3.5in height 0.4pt \color{black}\vspace*{5pt}} 
\setcounter{secnumdepth}{5}

\makeatletter 
\renewcommand\@biblabel[1]{#1}            
\renewcommand\@makefntext[1]%
{\noindent\makebox[0pt][r]{\@thefnmark\,}#1}
\makeatother 
\renewcommand{\figurename}{\small{Fig.}~}
\sectionfont{\sffamily\Large}
\subsectionfont{\normalsize}
\subsubsectionfont{\bf}
\setstretch{1.125} 
\setlength{\skip\footins}{0.8cm}
\setlength{\footnotesep}{0.25cm}
\setlength{\jot}{10pt}
\titlespacing*{\section}{0pt}{4pt}{4pt}
\titlespacing*{\subsection}{0pt}{15pt}{1pt}

\fancyfoot{}
\fancyfoot[LO,RE]{\vspace{-7.1pt}\includegraphics[height=9pt]{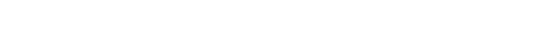}}
\fancyfoot[CO]{\vspace{-7.1pt}\hspace{13.2cm}\includegraphics{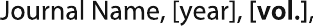}}
\fancyfoot[CE]{\vspace{-7.2pt}\hspace{-14.2cm}\includegraphics{head_foot/RF}}
\fancyfoot[RO]{\footnotesize{\sffamily{1--\pageref{LastPage} ~\textbar  \hspace{2pt}\thepage}}}
\fancyfoot[LE]{\footnotesize{\sffamily{\thepage~\textbar\hspace{3.45cm} 1--\pageref{LastPage}}}}
\fancyhead{}
\renewcommand{\headrulewidth}{0pt} 
\renewcommand{\footrulewidth}{0pt}
\setlength{\arrayrulewidth}{1pt}
\setlength{\columnsep}{6.5mm}
\setlength\bibsep{1pt}

\makeatletter 
\newlength{\figrulesep} 
\setlength{\figrulesep}{0.5\textfloatsep} 

\newcommand{\topfigrule}{\vspace*{-1pt}%
\noindent{\color{cream}\rule[-\figrulesep]{\columnwidth}{1.5pt}} }

\newcommand{\botfigrule}{\vspace*{-2pt}%
\noindent{\color{cream}\rule[\figrulesep]{\columnwidth}{1.5pt}} }

\newcommand{\dblfigrule}{\vspace*{-1pt}%
\noindent{\color{cream}\rule[-\figrulesep]{\textwidth}{1.5pt}} }

\makeatother

\twocolumn[
  \begin{@twocolumnfalse}
\vspace{3cm}
\sffamily
\begin{tabular}{m{4.5cm} p{13.5cm} }

\includegraphics{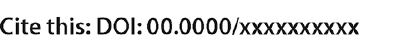} & \noindent\LARGE{\textbf{Full-Dimensional Schr\"odinger Wavefunction Calculations using
Tensors and Quantum Computers: the Cartesian component-separated approach}} \\
\vspace{0.3cm} & \vspace{0.3cm} \\

 & \noindent\large{Bill Poirier$^{\ast}$\textit{$^{a}$} and Jonathan Jerke\textit{$^{b}$}} \\

\includegraphics{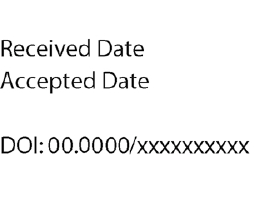} & \noindent\normalsize{Traditional methods in quantum chemistry rely on 
Hartree-Fock-based Slater-determinant (SD) representations, whose underlying zeroth-order picture assumes  separability 
by \emph{particle}.   Here, we explore a radically different  approach, based on separability by \emph{Cartesian component}, 
rather than by particle [\textit{J. Chem. Phys.}, 2018, \textbf{148}, 104101]. The approach appears to be very well suited for 
3D grid-based methods in quantum chemistry, and thereby also for so-called ``first-quantized'' quantum computing.  
We first present an overview of the approach as  implemented on classical computers, including numerical results that 
justify performance claims.  In particular, we perform numerical calculations with four explicit electrons that are 
equivalent to full-CI matrix diagonalization with nearly $10^{15}$ SDs.
 We then present an implementation for quantum computers, for which both the number of qubits, 
and the number of quantum gates, may be substantially reduced in comparison with other quantum circuitry that has been 
envisioned for implementing first-quantized ``quantum computational chemistry'' (QCC).} 
\end{tabular}

 \end{@twocolumnfalse} \vspace{0.6cm}

  ]

\renewcommand*\rmdefault{bch}\normalfont\upshape
\rmfamily
\section*{}
\vspace{-1cm}


\footnotetext{\textit{$^{a}$~Department of Chemistry and Biochemistry and Department of Physics, Texas Tech University, P.O. Box 41061, Lubbock, Texas 79409-1061, USA. Fax: 806-742-1289; Tel: 806-834-3099; E-mail: Bill.Poirier@ttu.edu}}
\footnotetext{\textit{$^{b}$~Email: jonathanjerke@aya.yale.edu }}






\section{Introduction}
\label{intro}

The importance of quantum chemistry simulations\cite{szabo,jensen,helgaker,RN544} across a  broad swathe of science and engineering 
lies beyond question.  In  the field of \emph{quantum computational chemistry} (QCC)---i.e., quantum chemistry run on quantum computers\cite{poplavskii75,feynman82,lloyd96,abrams97,zalka98,lidar99,abrams99,nielsen,aspuru05,kassal08,whitfield11,brown10,christiansen12,georgescu14,kais,huh15,babbush15,kivlichan17,babbush17,babbush18,babbush18b,babbush19,low19,kivlichan19,izmaylov19,parrish19,altman19,cao19,alexeev19,bauer20,aspuru20}---most  efforts to date have attempted to simply  “translate” as much as possible of the standard  electronic structure 
methodology, developed over many decades for classical computers. Given the vast
and fundamental differences between quantum and classical  architectures, however,  this may not be the most 
profitable or natural approach.

In particular, standard  methods (what in the QCC parlance are often called “second-quantized,” in a sense to be 
defined shortly) rely on Hartree-Fock (HF) (or other) Slater-determinant (SD) basis set
representations.\cite{szabo,jensen,helgaker,lloyd96,abrams97,low19,aspuru20}  
The underlying zeroth-order picture assumes separable products of single-particle molecular (spin)-orbital functions. 
Depending on the further approximations used by the particular implementation, standard methods are:
\begin{itemize}
\item[(a)]   not always reliably accurate for strongly correlated systems (except in a full-CI matrix calculation that can be
difficult to achieve in practice).
\item[(b)]     generally not designed to compute many accurate excited electronic states (i.e., tens or hundreds of states, 
including position-dependent wavefunctions).
\item[(c)]   not directly applicable to the combined electron-nuclear motion quantum many-body problem
(describing the total molecule).
\item[(d)]    not generally regarded as the most viable candidates for long-term, full-fledged  quantum computing simulation.
\end{itemize}
In the QCC context, standard second-order methods are   being pursued as an important  but near-term strategy---e.g.,
for use on hybrid quantum/classical computing platforms.\cite{babbush18,babbush18b,kivlichan19,izmaylov19,aspuru20}

In the long term, the most effective QCC algorithms are likely to be \emph{first-quantized} 
strategies\cite{abrams97,zalka98,lidar99,abrams99,nielsen,aspuru05,kassal08,whitfield11,huh15,babbush15,babbush17,kivlichan17,babbush19,aspuru20}—i.e., those that rely on explicit coordinate-grid-based representations of the entire electronic (or electron-nuclear) 
wavefunction, across all system coordinates.  Note that we are using ``first-'' and ``second-quantized''  in the QCC 
sense,\cite{lloyd96,abrams97,low19,aspuru20}  which differs slightly from standard electronic structure 
usage,\cite{szabo,jensen,helgaker}  as has caused some confusion in the past. To make matters worse, in  the QCC context, 
the term ``first-quantized'' can also be used to describe plane-wave and  more sophisticated basis set representations, 
similar to the molecular orbitals of the second-quantized approach. However, insofar as QCC and this paper are concerned,
\emph{the defining feature of any ``second-quantized'' calculation is its reliance on an SD basis representation.} 
First-quantized methods are not constrained in this manner, and thereby not necessarily subject to the limitations
 (a) through (d) above. 
This issue is taken up again in Sec.~\ref{methodology}.

In comparison with second-quantized methods, the chief drawbacks of first-quantized methods are that: (i) \emph{much} larger basis
set sizes are typically needed to represent the electronic wavefunction (at least for grid-based representations), and; (ii) 
antisymmetry is not ``built in'' from the start, but most be explicitly imposed. 
These present very substantial obstacles on classical computers, so much so such that to date, only a few
serious attempts have been made to develop viable first-quantized electronic  structure methods (mostly in the context of 
two-electron basis functions or geminals).\cite{lowdin56,silver69,gonis98,mazziotti00,mazziotti10}  The primary reason is the 
implied exponential scaling of computational resources with respect to the number of particles ($N$), or system dimensionality ($3N$). 

On true quantum computers, however---at least as they are currently envisioned to exist within a few years---neither 
drawback (i) nor (ii) above is  expected to be especially daunting. With regard to (i),  the exponential scaling becomes  
\emph{linear} scaling---at least in terms of the total number of logical qubits required to store the  wavefunction 
(the algorithmic scaling is at worst polynomial in $N$).  More specifically, storing the wavefunction will require
something like $3Nn$ logical qubits (ignoring ancilla qubits, error correction qubits, etc.), with a minimum of $n \approx 7$. 
As for (ii), this is taken up in Sec.~\ref{antisymmetry}.
All in all, first-quantized circuit-based QCC methods are regarded to be more likely than second-quantized circuits to achieve 
``quantum supremacy''\cite{aspuru05,georgescu14,altman19,aspuru20}---by which is meant, a calculation that would be
impossible using the most advanced classical computers.  

It has also been argued that first-quantized calculations are more accurate than second-quantized (apart from full-CI) calculations, 
as fewer approximations and assumptions are relied upon.\cite{alexeev19,bauer20,aspuru20}  Though likely largely true, 
this author is not fully convinced by such arguments; in particular, grid-based calculations can introduce 
errors of their own---e.g., if  a quadrature approximation  is used.\cite{RN553,colbert92,szalay96,light00,poirier02dvrlj,littlejohn02b,jerke15}
On the other hand, it is certainly true that removing the SD constraint from the basis set opens the door to new types of highly 
efficient representations for handling strongly correlated systems, as deserves future exploration (Sec.~\ref{antisymmetry}).

First-quantized methods also enable direct treatment of the combined 
electron-nuclear motion quantum many-body problem---which is simply not possible in a second-quantized SD
treatment.\cite{aspuru20}  This feature turns out to be \emph{vitally} important in the QCC context, 
owing to the fact that separate electronic and nuclear motion calculations require  large-dimensional potential energy surfaces,
whose explicit calculation itself scales exponentially---even on a quantum computer.\cite{kassal08,christiansen12,aspuru20}  
In contrast, a direct, first-quantized QCC calculation avoids this difficulty by solving the combined electron-nuclear problem 
``all at once.'' 

On the other hand, the tremendous promise of first-quantized QCC is  mitigated by the limits of present-day quantum 
hardware, which  can accommodate no more than $\sim$100 or so qubits in total.\cite{aspuru20}
 Of even greater concern is the limited number 
of quantum gates that can be reliably implemented within a  quantum circuit---given current gate fidelities, and 
without viable error-correction or fault-tolerant 
strategies yet in play.\cite{ogorman17,preskill18,babbush18b,kivlichan19,low19,aspuru20}  
For realistic QCC applications, the  circuit complexities required by first-quantized methods are still 
beyond what can be accommodated in the
near future. Hence,  first-quantized QCC is considered a ``long-term''  strategy, receiving far less attention to date than has
second-quantized QCC.\cite{babbush15,babbush18,aspuru20}  Nevertheless 
by developing better first-quantized  
QCC algorithms today---i.e., with improved scaling, higher accuracy, fewer required qubits and quantum gates, etc.---we
we will arrive that much sooner at the desired tomorrow.

Recent efforts 
 have concentrated on improving 
 scaling and scalability,\cite{kivlichan17,babbush18,babbush18b,babbush19,kivlichan19,aspuru20} 
and
on efficient  strategies for implementing the 
requisite unitary  evolutions [e.g.,  linear combination of unitary 
(LCU) operators].\cite{berry14,low16,babbush19,berry19b,aspuru20}
Additionally, 
 Galerkin discretization offers a rigorously variational approach with respect to basis size,\cite{babbush18,babbush19,aspuru20}
 by eliminating all quadrature error. Such techniques represent the 
current state of the art in first-quantized QCC development, even when grid or plane-wave representations  are 
used (second-quantized QCC has met with more success, to date, in terms of specialized orbitals/basis 
sets\cite{whitfield11,berry19,berry19b,low19,aspuru20,babbushcomm}). 

The present work approaches first-quantized QCC from a 
 complementary direction---by exploiting
alternate representations that may require fewer qubits, and may also give rise to quantum circuits with fewer 
quantum gates.  Specifically, we employ a 
radically different, tensor-product representation of the exact Coulomb potential energy operators, based on separability 
by \emph{Cartesian component} rather than by particle.\cite{jerke15,jerke18,jerke19}   The resultant ``Cartesian 
component-separated'' (CCS) approach is necessarily first-quantized (by the definition given above), and 
otherwise turns out to resolve all of the limitations [(a)--(d) as listed above] of the standard second-quantized SD methods.

On classical computers, the CCS tensor-product representation of the Hamiltonian  may be
combined with a similar tensor-product form for the wavefunction---giving rise to a highly efficient 
computational quantum chemistry scheme, exhibiting many orders-of-magnitude reduction in numerical effort,
as compared to explicit representations.
Using only grid or plane-wave basis sets, classical CCS calculations have been performed to compute multiple excited electronic 
states (including wavefunctions) for atoms up to lithium,\cite{jerke18,coulson61}  for other central-force applications 
with non-Coulomb interactions,\cite{jerke19} and for strongly correlated electron gases with up to $N=4$ explicit 
electrons.\cite{ceperley80,esteban12,varas16,bittner}  
The method is also currently being applied to solve the combined electron-nuclear motion problem (for H$_2^+$\cite{perez13} and 
H$_2$\cite{liu09})---which has been identified as being important for long-term QCC, as discussed.\cite{kassal08,christiansen12,aspuru20} 

To further demonstrate the feasibility of the classical CCS approach, we have for this paper performed an
explicit $N=4$ calculation of the ``harmonium'' central-force system,\cite{RN756,RN773,RN755,jerke19}
 using only a plane-wave representation. 
The underlying basis size required is astronomical---i.e.,   $N_B \approx 10^{15}$.  This calculation may be 
compared with a full configuration interaction (full-CI) matrix diagonalization calculation conducted using 
a similar number of SD basis functions, and with $M \approx 5000$ single-particle (spin-)orbitals.  

Such classical CCS calculations are important not only in their own right,
but also to establish accurate benchmarks for future first-quantized QCC calculations---particularly those using 
grid or plane-wave basis representations.   On the other hand, the classical CCS calculations conducted to date
may all be regarded as ``preliminary,'' in the sense that they have all used a very poor choice of basis that does not exploit 
CCS's primary advantages vis-\`a-vis correlation.   In this paper, for the first time, we set the theoretical stage
for ``optimized'' CCS representations (analogous to Hartree-Fock) that take maximum advantage of correlation, and
may thus give rise to orders-of-magnitude reductions in the \emph{underlying} 
basis size, $N_B$---irrespective of the subsequent tensor-product reductions that classical CCS also provides.   
Such developments are expected to greatly improve what classical CCS may ultimately be able to do---with 
respect to system size $N$, overall numerical accuracy,  or both. 

The remainder of this paper is  as follows. In Sec.~\ref{methodology}, we lay out the advantages
and operation of the first-quantized CCS approach in more detail, discussing also the form of the tensor-product
representations used for both the Hamiltonian operators and the wavefunctions. This section concludes
with a discussion of permutation symmetry, and how to rigorously enforce the condition of antisymmetry. 
Sec.~\ref{results} analyzes  various classical CCS results that have been obtained to date (including 
the brand new results for $N=4$ harmonium), with an eye towards what these  imply about the likely
capabilities of the CCS approach in future, both in the classical and quantum computing contexts. 
Sec.~\ref{quantum} then lays out our proposed quantum circuitry for implementing CCS as a 
QCC method---encompassing 
both time-independent eigenstate solutions [via quantum eigensolvers such as quantum phase 
estimation (QPE)\cite{abrams99,nielsen,parrish19,izmaylov19,aspuru20}], 
as well as time-dependent solutions. 
 
Although it is highly likely that a CCS QCC approach could be applied in conjunction with more 
sophisticated unitary evolution strategies such as LCU,\cite{babbushcomm} for pedagogical simplicity,  we address here 
only the more standard and straightforward ``canonical Trotter method.''\cite{trotter59,zalka98,nielsen,georgescu14,kivlichan19,aspuru20} 
In this context, we can likely realize a  reduction in  the number of both qubits and quantum gates,
in comparison with other first-quantized quantum circuitry that has been envisioned for QCC.
Two alternate CCS implementations are presented, which can provide a potentially useful ``engineering tradeoff'' 
between quantum circuitry bottlenecks. 
Finally, future prospects are discussed, which turn out to hinge on (of all things) an efficient quantum  implementation
of the exponential function. 


\section{Methodology}
\label{methodology}

\subsection{Advantages of First Quantization and the CCS Approach}
\label{advantages}

In order to better understand the ``first-'' vs. ``second-quantized'' terminology in the  context of QCC, it is helpful
to consider the corresponding quantum \emph{operators}, as distinct from their basis set representations. In the
traditional first-quantized formalism, the Hamiltonian operator is expressed using electron position coordinates 
as indicated in \eq{Ham}.\cite{szabo,jensen,helgaker}  
Of course for numerical purposes, this operator may be represented as a matrix using \emph{any} desired
representational basis set---i.e., any set of orthonormal functions on the entire $3N$-dimensional 
``configuration space.'' (Note that the latter phrase is being used in the generic sense to mean the space of all particle 
positions, rather than in the electronic structure sense of orbital occupations).  

In particular, the basis  functions need \emph{not} be SDs. One obvious choice of
basis that is decidedly not of  the SD form is suggested by \eq{Ham} itself---i.e., the position eigenstates. 
In principle, these are Dirac delta functions; in practice, due to basis set truncation,
these are \emph{sinc  functions},\cite{RN553,colbert92,szalay96,light00,poirier02dvrlj,littlejohn02b,jerke15}
associated  with a uniform lattice of grid points, and spanning the same Hilbert subspace as a
corresponding  band-limited set of plane waves. Both sinc (aka ``plane-wave dual'') and plane-wave
basis sets have been 
considered in the QCC context,\cite{aspuru05,babbush18,babbush19,aspuru20} 
and have also been employed with the CCS approach.  However, we again emphasize that CCS offers
tremendous flexibility in general, and  a great many other choices of CCS basis set are also possible---including those for 
which substantial particle correlation is  ``built in'' to the individual basis functions (Sec.~\ref{antisymmetry}).

What is not built in to any first-quantized approach, however, is antisymmetry---which, as noted, must therefore be
explicitly imposed.  Here, the second-quantized approach has the upper hand, essentially  because it treats
operators and basis functions on an equal footing.\cite{helgaker}  In particular, second-quantized
Hamiltonian operators are sums of
products of creation and annihilation operators, acting on a set of $M \ge N$ single-particle (spin-)orbitals. 
Second quantized basis functions 
are  ``occupation number'' (ON) vectors---i.e., binary strings denoting which of the $M$ orbitals are
occupied---spanning the $2^M$-dimensional Fock space (in the ``grand canonical'' sense where $N$ may vary,
although for all of our applications, $N$ is fixed).  All ON vectors may be 
obtained by applying orbital-specific creation operators to the ``vacuum state'' of no electrons.  
The second-quantized approach is thus  more abstract and algebraic than the first-quantized approach,
although at the end of the day, 
\emph{ON vectors always correspond to properly antisymmetrized SDs of single-particle orbitals}.  
This is both an advantage (because antisymmetrization need not be dealt with explicitly) and a disadvantage
(because  \emph{no other} type of basis may be considered). 

The CCS approach as advocated here does not rely on SDs, single-particle orbitals, nor particle
separability of any kind.  It is therefore very much a first-quantized approach, by any reckoning. Using
the representational basis, the Hamiltonian operator is (implicitly) represented as an  $N_B \times N_B$ 
matrix, which is then diagonalized. In principle, a large number of accurately converged eigenstates may be
obtained, with rigorous variational convergence to the exact results guaranteed in the complete basis set (CBS) 
limit.  The approach may therefore be compared with full-CI matrix diagonalization calculations from the 
second-quantized realm [i.e., for which all $N_B=\of{M \atop N}$ possible SDs are retained].  However,
CCS calculations are \emph{much} more flexible than full CI---not only with respect to the choice of basis itself,
but also in terms of how the basis is truncated.  More specifically, in a full-CI calculation, $M$  is the only  
convergence parameter (with $M \rightarrow \infty$ defining  the CBS limit). 
In contrast, in the CCS case, there are up to 3$N$ (though usually just 3) 
separate basis  truncation parameters,  each of which can be adjusted 
independently for maximum efficiency.  

There is one more advantage of the CCS approach that is particular relevant, especially
for QCC.  Since it is not limited to SDs, it can be applied directly to the combined electron-nuclear 
motion quantum many-body problem.  Ordinarily, one imposes the Born-Oppenheimer (BO) 
approximation,\cite{szabo,jensen} to
separate the electronic and nuclear motion problems, although this introduces a source of error.  More accurate 
results may therefore be obtained by solving the combined problem directly.  In the QCC context however, there is a  
much  more compelling and practical reason for wanting to solve the combined problem. 
This is because the BO approximation necessitates the development of an explicit potential energy surface (PES)
over the entire 3$N$-dimensional space---a prospect which even on a quantum computer, scales exponentially
with $N$.\cite{kassal08,christiansen12,aspuru20}  By solving the combined electron-nuclear motion problem 
``all at once,''  such PES problems are completely avoided---as are all BO-type errors. 

The above considerations justify claims (a) thru (d) in Sec.~\ref{intro} (although additional
arguments  are also provided throughout this paper). Going forward, we shall discuss both the CCS formalism 
itself, as well as its application to first-quantized QCC.  Since both topics are likely unfamiliar to many readers, 
it is pedagogically useful to ``divorce'' them from each other.  Accordingly, 
in this section we focus exclusively on the former, presenting an overview of how  classical CCS  is 
implemented, and explaining how it relates to---but differs from---conventional 
quantum chemistry methods.  The QCC case  is then taken up again in Sec.~\ref{quantum}.

In particular, we present here the most \emph{generalized} version of the CCS theory---i.e., which does \emph{not} presume 
 separability by particle, and
therefore allows for substantial correlation to be built directly 
into the  representational basis itself. 
After all,
this is the most important and distinctive feature of the CCS approach.
On the other hand, all classical CCS 
calculations performed to date\cite{jerke15,jerke18,jerke19,bittner} have used sinc or plane wave basis sets, 
both of which are separable with respect to \emph{both} particle \emph{and} Cartesian component. 
Such basis sets have provided a natural and straightforward starting-off point for  classical computations, 
and will also form the basis of our (initial) 
QCC algorithm in Sec.~\ref{quantum}.  On the other hand, the past emphasis on plane-wave---or
at least particle-separated---basis sets may have also served to obscure the greater significance of CCS.   That
situation will be rectified here.

\subsection{CCS  Explicit and Tensor SOP Representations of the Wavefunction}
\label{CCSwaverep}

For a system of $N$ electrons, let $\vec r_i = (x_i,y_i,z_i)$ represent the Cartesian components of the $i$'th electron
position.  We shall work with a general CCS representational basis of the form 
\eb
	\Phi_{\vec m}(\vec r_1,\ldots,\vec r_N) = X_{m_x}(x_1,\ldots,x_N) Y_{m_y}(y_1,\ldots,y_N) Z_{m_z}(z_1,\ldots,z_N),
	\label{CCSbasis}
\ee
where $\vec m = (m_x,m_y,m_z)$.  Note that the basis spans the entire $3N$-dimensional configuration space. 
  The form of \eq{CCSbasis} should be contrasted with  the standard SD form. 
  Apart from antisymmetrization, SDs  are separable by \emph{particle}, 
whereas \eq{CCSbasis} is separable only by \emph{Cartesian component}.  This means that full correlation across (in principle) 
all $N$ electrons can be built directly into the  basis representation.   On the other hand, \eq{CCSbasis} does not (yet)
incorporate either permutation symmetry or spin.  

The total size of the representational basis of \eq{CCSbasis} is $N_B=M_x M_y M_z$, where $M_d$ is the basis size
for the $d$'th Cartesian component (i.e., $d=\{x,y,z\}$), so that $1 \le m_d \le M_d$.  Alternatively, $M_d$ may be
thought of as the number of CCS ``orbitals'' associated with the $d$'th Cartesian component---which need not be the same
for $x$, $y$, and $z$.    
Obviously, $M_d$ increases
quickly with $N$, even if an excellent, highly efficient basis is chosen.  For the (typically) worst-case choice of a plane wave 
(or sinc) basis, we have (e.g., for the $x$ component),
\eb
	X_{m_x}(x_1,\ldots,x_N) = \phi^{x}_{m_{1x}}(x_1) \cdots \phi^{x}_{m_{Nx}}(x_N),
	\label{planebasis}
\ee
with $m_x = (m_{1x}, \ldots m_{Nx})$, etc. 
 If  $1 \le m_{id} \le L$, so that there are $L$ basis functions per Cartesian 
component \emph{per particle}, then the number of CCS orbitals per component is $M_d = L^N$,
and the total  basis size is $N_B= L^{3N}$.

For molecular applications treating core electrons exactly, $L\approx 100$ might be  a typical value. 
In comparison with the second-quantized SD approach, this would  correspond to 
$M = L^3 \approx 10^6$  single-particle orbitals---a ``worst-case''  choice of basis indeed!  
Recognizing further that explicit matrix  representations 
would require storage of $N_B^2 = L^{6N}$ elements in all, it becomes clear that even two-electron 
applications lie beyond what would be possible on any  classical computer.
Use of smarter CCS basis sets, \emph{not} of the \eq{planebasis} form, can of course greatly reduce the requisite 
$M_d$ and $N_B$ sizes, thereby effectively increasing the number of electrons that might be treated explicitly.
In any case, it is  clear  that the scaling of explicit vector and matrix
representations  with respect to $N$  poses a severe challenge for classical computers.

It is at this juncture, then,  that the classical and quantum computational  strategies  part company---at least with
respect to how the $N$-electron CCS wavefunction is actually represented numerically.  On the quantum  side, 
the exponential growth of $N_B$ with $N$ does not pose a fundamental problem; as discussed, this 
corresponds to \emph{linear} growth in the  number of qubits (Sec.~\ref{quantum}). 
We therefore simply use the explicit vector representation of the $N$-electron wavefunction 
 implied by  \eq{CCSbasis}, i.e.,
\eb
	\Psi(\vec r_1,\ldots,\vec r_N) \approx \sum_{m_x=1}^{M_x} \sum_{m_y=1}^{M_y} \sum_{m_z=1}^{M_z}
	\Psi_{m_x m_y m_z} \Phi_{\vec m}(\vec r_1,\ldots,\vec r_N), 
	\label{directpsi}
\ee 
[together with \eq{planebasis}], as is. More  specifically,  in the present QCC CCS implementation, 
both plane wave and sinc representations 
are used, with explicit storage of the vector, $\Psi_{m_x m_y m_z}$, requiring $N_B=L^{3N}$ elements, as discussed.  
[As per Sec.~\ref{quantum}, the quantum Fourier transform\cite{shor94,abrams99,nielsen} (QFT) is used to switch 
between these two representations.]   

In contrast, the classical CCS implementation replaces the explicit vector representation 
of \eq{directpsi} with the following [generic \eq{CCSbasis} basis] tensor ``sum of product'' (SOP) form:
 \eb
        \Psi_\Lambda(\vec r_1,\ldots,\vec r_N) \approx \sum_{\lambda=1}^\Lambda 
        \sum_{m_x=1}^{M_x} \sum_{m_y=1}^{M_y} \sum_{m_z=1}^{M_z}
        X^\lambda_{m_x}  Y^\lambda_{m_y}  Z^\lambda_{m_z}  \Phi_{\vec m}(\vec r_1,\ldots,\vec r_N)
        \label{tensorpsi}
\ee
 In \eq{tensorpsi}, the number of elements to be stored is now only $\Lambda (M_x+M_y+M_z)$---which, in 
 general, is only a tiny fraction of $N_B$.  Of course---as with all tensor methods---this requires that a 
 a sufficiently good wavefunction approximation is achievable using a reasonably small number of 
 terms, $\Lambda$. 
 For all CCS applications to date, this has been achieved with $\Lambda$ on the order of 100.\cite{jerke15,jerke18,jerke19,bittner}
 Consequently, explicit $N$-electron wavefunctions have been stored using only a modest amount of RAM
 on classical computers---even up to $N=4$.   
 
 Such enormous reductions are not atypical when tensor SOPs are employed.
 Recent years have seen an explosion in the use of modern tensor methods across a vast range of data science
 applications, as a tool for drastically reducing data storage requirements.\cite{beylkin05,khoromskaia15,leclerc16}  
 Even in the electronic structure
 realm, use of tensor SOPs to represent the electronic state is not new, and has been applied to great effect for a 
 number of years now.\cite{hohenstein12,parrish13,shepard14,shepard19,lischka20}   
 Nevertheless, our use of tensor SOPs is radically different from earlier efforts,
 in two important ways.  To begin with, it is applied across a CCS basis that is in general separable only by Cartesian 
 component [\eq{CCSbasis}] and not by particle.  The very nature of the tensor products involved is thus quite
 different from previous efforts based on particle-separated SD orbitals. 
 
 Secondly---and arguably more importantly---the CCS approach enables a novel tensor SOP representation of the 
 exact, full $N$-electron Hamiltonian, that is remarkably efficient and accurate, and that can handle  Coulombic
 singularities with ease---even in a sinc grid-based representation.  This is because potential energy functions
 are not simply evaluated at the grid points.   As a consequence, it is possible to  place a grid point even directly 
 on a Coulombic singularity, with no particularly damaging numerical repercussions.\cite{jerke15,jerke18}
 Likewise---and regardless of the particular CCS basis representation---the Hamiltonian tensor SOP is not 
 some arbitrary numerical approximation to the explicit matrix representation, but is instead a  fundamental identity that arises 
 naturally from a CCS analysis of the underlying operators.  This is described in detail in the following subsection.  
 
\subsection{CCS Tensor SOP Representation of the Hamiltonian}
\label{CCSHamrep}

The first-quantized electronic structure Hamiltonian $\hat H$ is generally of the form
\ea{
	\hat H  & =  & \sum_{i=1}^N \of{ {{\hat p}^2_{ix} \over 2 m_e} + {{\hat p}^2_{iy} \over 2 m_e} + {{\hat p}^2_{iz} \over 2 m_e} } 
	                           + \nonumber \\
	 & &  \sum_{i=1}^N V_{{\rm ext}}(x_i, y_i, z_i) +\sum_{i<j}^N V_{ee}(x_i, y_i, z_i, x_j, y_j, z_j),  \label{Ham}
}
where $m_e$ is the mass of the electron.
The first sum in \eq{Ham} above represents the kinetic energy contribution, which separates by both particle and
Cartesian component.
 The remaining,  external ($V_{{\rm ext}}$) and pair repulsion ($V_{ee}$) potential energy terms, respectively, involve
 three or six coordinates each.  These are generally taken to have Coulombic form, although the present method is not 
 necessarily restricted to Hamiltonians of this kind.  
 
 Assuming a Coulombic form for the electron pair repulsion contribution, we have
\ea{
V_{ee}(x_i, y_i, z_i, x_j, y_j, z_j) & = & V_{ee}(s) = 1/s^{1/2}, \quad \mbox{with} \\
s & = & (x_i-x_j)^2 + (y_i-y_j)^2 + (z_i-z_j)^2. \nonumber 
}
Inverse Laplace transformation then yields the following exact form:
\ea{
	\hat V_{ee} & =  & { 2 \over \sqrt{\pi}}  \int_0^\infty  \hat \gamma_x (\beta) \hat\gamma_y (\beta) \hat\gamma_z (\beta) \, d\beta ,
	\quad \mbox{where} \label{betaint} \\
	\hat \gamma_d(\beta)  & = & \exp\!{\sof{-\beta^2 (d_i-d_j)^2}} 
}
In \eq{betaint} above, the individual $\hat \gamma_d$ factors in the integrand 
(e.g., $\hat \gamma_x(\beta) = \exp\!{\sof{-\beta^2 (x_i-x_j)^2}}$) 
are two-particle operators involving just a single Cartesian component, $d$. The critical point is that 
{\em these 
factors separate by \emph{Cartesian component}, and \emph{not} by particle}.   Note that spherical symmetry implies the 
same form for  all of the $\hat \gamma_d(\beta)$, and so in practice the $d$ subscript can sometimes be dropped. 

A form similar to \eq{betaint} has been used previously in electronic structure, in the context of Rys polynomial
integrals.\cite{rys83} 
In that context, the component separability of \eq{betaint} is exploited in order to simplify the evaluation of Coulomb and 
exchange integrals---a perennial, if well-established challenge in the electronic structure community. 
Here, however, the Cartesian component separability of the $\beta$ integrand is used in an 
entirely different manner---i.e., to construct a natural tensor SOP representation of the two-electron pair repulsion
operator itself.  Also, Rys considered only Gaussian basis functions, i.e. not general CCS basis representations
of the form of \eq{directpsi}, and certainly not the tensor-SOP reduction [i.e., \eq{tensorpsi}]. 

To convert \eq{betaint} above into an explicitly tensor-SOP form, the  one-dimensional integral over $\beta$ is evaluated 
 numerically by quadrature, on a finite and discrete set of  quadrature points, $\beta_\lambda$. 
The integral is then effectively replaced with a tensor-product sum of the form
\eb
	\hat V_{ee} = \sum_{\lambda=1}^{\Lambda_{ee}} w_\lambda 
	{\hat \gamma}_x^\lambda  {\hat \gamma}_y^\lambda  {\hat \gamma}_z^\lambda,  \label{gammasum}
\ee
where the $w_\lambda$ are the quadrature weights [into which the $2/\sqrt{\pi}$ factor from \eq{betaint} 
has also been subsumed].
Remarkably, quadrature schemes with as few as $\Lambda_{ee}=22$ points have been found that
 provide accuracies near to
machine precision, with substantially smaller $\Lambda_{ee}$ values required for chemical accuracy
(i.e., $\sim$1--2 millihartree).
To our knowledge, no particle-separated tensor-SOP matrix representations of $\hat V_{ee}$  provide 
comparable accuracy with as few terms.\cite{hohenstein12,parrish13} That said, recent graphically contracted
function (GCF) and all configuration mean energy (ACME) methods show that such dramatic reductions 
are possible, at least in the multiconfiguration self-consistent field (MCSCF, including complete active space or CASSCF) 
and state-averaged contexts.\cite{shepard14,shepard19,lischka20}

The individual factors in \eq{gammasum} are two-particle single-component ``potential energy'' operators that 
get represented as single-component matrices as follows (e.g., for $x$):
\ea{
\gamma^\lambda_{n_x m_x}  & = & (\beta_\lambda^2 \mid\mid n_x, m_x)   \label{gammamat} \\
   & = &  \int  X_{n_x}^*(x_1,\ldots,x_N) e^{-\beta_\lambda^2 (x_i-x_j)^2}   X_{m_x}(x_1,\ldots,x_N)
\, dx_1 \ldots dx_N \nonumber
}
For  basis functions that are also particle-separated as per \eq{planebasis} (e.g., plane waves), 
the integrals in \eq{gammamat} reduce to two dimensions. Still simpler forms will be considered in Sec.~\ref{quantum},
e.g. based on quadrature.  Even for optimized, highly correlated CCS basis functions of the
\eq{CCSbasis} form,  the integrals of \eq{gammamat} can be efficiently integrated numerically---e.g., using 
a further tensor-SOP decomposition.  

In any event, these CCS Coulomb integrals need only be performed
once for all time, for a given choice of basis---irrespective of any subsequent electronic structure calculations 
to which they might be applied. Once the component tensor matrices have been determined as per above, 
they are used to construct a tensor-SOP representation of the electron pair repulsion  contribution
to the Hamiltonian operator that requires storage of at most  $\Lambda_{ee} (M_x^2+M_y^2+M_z^2)$ elements.  
For plane-wave and other particle-separated CCS basis sets of the \eq{planebasis} form, only $\Lambda_{ee} L^4$ 
elements need be stored.

There remain, in the Hamiltonian of \eq{Ham}, additional contributions from the kinetic energy, and the external potential
energy, $V_{{\rm ext}}$.  If the latter represents Coulombic attraction to nuclei---as is expected to be the most common
case---then a CCS tensor SOP form very similar to \eq{betaint} may be employed.\cite{jerke15,jerke18,jerke19}  
Matrix elements are also defined similarly to \eq{gammamat}.  
Storage requires $\Lambda_{Ze} (M_x^2+M_y^2+M_z^2)$ elements in the most general
case---with $\Lambda_{Ze}$ about as small as $\Lambda_{ee}$, despite the fact that the Coulomb interaction is now
attractive.  For CCS basis sets that are in addition particle-separated [\eq{planebasis}], the scaling is only $\Lambda_{Ze} L^2$. 

On the other hand, it must be recognized that attractive, bare Coulomb potentials tend to generate large
errors when plane-wave or sinc basis representations are used.  Stated differently, such basis sets
represent a very inefficient choice when $V_{{\rm ext}}(\vec r_i)  = -\sum_k Z_k/|\vec r_i -\vec R_k|$---for which 
$L$ values  in excess of 100 (or  equivalently, $M >  10^6$) may become required in order to attain chemical 
accuracy, as discussed.  
Better CCS basis sets, particularly 
highly correlated ones, will provide much better performance, in terms of either: (a) greatly reducing the overall basis size 
$N_B$ needed to achieve a given level of accuracy, or; (b)  greatly improving the numerical convergence accuracy
for a given $N_B$. This issue is taken up again in Secs.~\ref{antisymmetry} and~\ref{results}. 

It should also be stated that  Coulomb potentials are far from the only type for which a Laplace-transformed CCS 
SOP form can be obtained. In particular, Yukawa and long-range Ewald potentials have also been represented
in this fashion.\cite{jerke19}  There are also, of course, external potentials that naturally adhere to the CCS 
form---such as isotropic harmonic oscillator potentials (i.e. ``harmonium''), which have also been considered in this 
context\cite{RN756,RN773,RN755,jerke19}
(Sec.~\ref{results}). Finally,
we point out that the Cartesian kinetic energy operator itself [\eq{Ham}] is also naturally CCS, and therefore
also straightforward and inexpensive to represent in a CCS basis.  For further details, please consult
[\cite{jerke18}] and [\cite{jerke19}], and also Secs.~\ref{results} and~\ref{quantum}.    

The discussion presented in this subsection primarily refers to how the various components of the
Hamiltonian matrix of \eq{Ham} are represented in tensor-SOP form on a \emph{classical} computer.  In 
particular, use of CCS tensor-SOP representations very dramatically reduces Hamiltonian matrix 
storage down from the $N_B^2$ elements that would otherwise have to be stored in an explicit representation.
Similarly dramatic savings using a CCS tensor-SOP representation for the wavefunction vector then make
it feasible to perform explicit $N$-electron calculations on a classical computer---to chemical accuracy, 
even for strongly correlated systems, and for highly excited electronic states, including wavefunctions---at 
least up to $N=4$ or so.  This is the story, more or less, on the classical computing side.

For the quantum computing implementation, however, things are rather different. Whereas on a quantum 
computer, qubits are indeed used to represent the wavefunction (albeit explicitly, rather than in tensor-SOP
form), the Hamiltonian matrix elements \emph{per se} are \emph{not} stored.  Instead, the
action of the Hamiltonian on the wavefunction is simulated through the use of appropriate 
\emph{quantum circuitry}---i.e., a collection of quantum gates---applied to the set of logical qubits used to 
represent $\Psi$.  

It is worth reflecting on the fundamental differences between the basic QCC framework as described above,
and that of traditional linear algebra as implemented on classical computers.  In the latter context,
a (direct) matrix--vector product is pretty much always implemented numerically in the same way, 
regardless of the choice of basis used to represent the matrix.  Of course, the matrix itself must be
stored explicitly, in addition to the vector, and it is in these explicit representations that one encounters 
variations from one basis set to another.  In the QCC context, however, the Hamiltonian matrix is reflected in the 
form of the  quantum circuitry itself---which consequently changes, depending on the choice of basis set.
Even for a given basis, there can be many different strategies, all designed
to perform the same operation more or less, but using different quantum circuitry.  

In any event, the QCC algorithm proposed in this work (Sec.~\ref{quantum}) certainly 
differs from previous quantum circuitry designed to do the same thing. In large measure, this is 
because it is based on the CCS tensor-SOP representation of $\hat H$, as described in this 
subsection, but there are some other new wrinkles  as well.

\subsection{Spin, Permutation Symmetry, and Enforcement of Rigorous Antisymmetry}
\label{antisymmetry}

Due to the Pauli Exclusion Principle, the entire $N$-electron wavefunction (including spatial and spin components) 
must be antisymmetric with respect to exchange of any two electrons. More properly, the wavefunction must belong to the 
totally antisymmetric (A$_2$) irreducible representation (irrep) of the $N$-body permutation group, $S_N$.
In the standard particle-separated approach, SDs are used to
automatically enforce rigorous antisymmetry of the entire electronic wavefunction.  This generally requires 
that spin states are built into the definition of the single-electron states (spin-orbitals).   since our CCS methodology is not 
based on SDs, some other means must be adopted to enforce rigorous antisymmetry of the entire 
electronic wavefunction. 

Spatial-spin decompositions  are fundamental to a wavefunction approach to electronic 
structure. 
For this paper, we shall always assume that the entire $N$-electron wavefunction can be
written as the product of an overall spatial wavefunction times an overall spin state. 
This spatial-spin factorization is legitimate for the 
Hamiltonians considered here, which do not involve spin explicitly (although the methodology could certainly be generalized 
to accommodate spin-orbit coupling or other corrections stemming from the Dirac equation). As a consequence, spin never 
enters into our numerical calculations at all, so that only the spatial part of the electronic wavefunction is ever explicitly represented
(e.g.,  in Sec.~\ref{CCSwaverep}).  Each computed 
numerical energy level would therefore have a $2^N$-fold spin degeneracy, were it not for  antisymmetry.
 
Conversely, the spin factor plays a huge rule in determining which spatial wavefunctions can give rise to entire 
wavefunctions that are antisymmetric. More generally, spin is used to determine the spin degeneracies of the computed 
electronic states. This is achieved using group theory, as follows. First, for a given permutation group, $S_N$, the 
direct-product group (representing the entire wavefunction) is determined. Next, the $2^N$-dimensional spin representation 
is decomposed into its $S_N$ irreps. Based on available spin irreps and degeneracies, and the direct-product table, 
the corresponding allowed spatial irreps and entire wavefunction degeneracies are determined. 

In the two-electron case, this is rather trivial: spatially symmetric (A$_1$) solutions correspond to spin-singlet (A$_2$) states, 
and spatially antisymmetric (A$_2$) solutions to spin-triplet (A$_1$) states, as per 
the standard and familiar rules. Thus, numerically computed (spatial) states for both $S_2$ irreps are physically realized. 
For larger $N$, however, the situation becomes much more complicated, in accord with the corresponding permutation symmetry 
groups, $S_N$.  For one thing, not all of the $S_N$ irreps are physically realized for $N>2$ . Additionally, there are 
degenerate irreps that appear, with far less trivial representations in terms of the underlying product basis. The direct-product tables 
are also decidedly more complicated.  Nevertheless, the requisite group theory has  been worked out by us, for all cases
up to $N \le 6$. 

\begin{table}[h]
\small
  \caption{$S_N$ irrep decompositions of spin state representations for $N \le 6$ (Column III).  This implies 
  allowed (Column IV) and forbidden (Column V) spatial irreps, as indicated, in order to enforce antisymmetry 
  (i.e., A$_2$ character) on the entire spatial-spin wavefunction.} 
  \label{symtab}
  \begin{tabular*}{0.48\textwidth}{ccccc}
    \hline
    $N$ & $2^N$ & Spin irrep  & \multicolumn{2}{c}{Spatial irreps} \\
    \cline{4-5}
      &    &   decomposition  & allowed  & forbidden \\
    \hline
    2 & 4    &   3A$_1 \oplus$A$_2$  	& A$_1$, A$_2$ &  \\
    3 & 8    &   4A$_1 \oplus$2E          & A$_2$, E         & A$_1$ \\
    4 & 16  &   5A$_1 \oplus$E$\oplus$3T$_2$             & A$_2$, E, T$_1$ & A$_1$, T$_2$ \\
    5 & 32  &   6A$_1 \oplus$4G$_1 \oplus$2H$_1$     & A$_2$, G$_2$, H$_2$  &  A$_1$, G$_1$, H$_1$, I \\
    6 & 64  &   7A$_1 \oplus$5H$_1 \oplus$H$_2 \oplus$3L$_1$ &A$_2$, H$_3$, H$_4$, L$_2$   
                                                                                         &  A$_1$, H$_1$, H$_2$, L$_1$, \\
        &      &                                &                                  &        M$_1$, M$_2$, S \\                                                                         
    \hline
  \end{tabular*}
\end{table}

Let us begin with the $S_N$ irrep decompositions of the $2^N$-dimensional spin state representations. These are 
summarized in Table~\ref{symtab}, Column III.  Note that A$_2$ never appears in the spin state irrep decompositions,
except for $N=2$.  This implies, e.g., that the A$_1$ spatial states are never physically realized for $N>2$.  More generally,
from the direct product tables for $S_N$, we have ascertained exactly which spatial irreps can give rise to an overall
A$_2$ state.  The resultant physically realizable spatial irreps, and also the forbidden  irreps, are listed in Columns IV and V, 
respectively, of Table~\ref{symtab}.  Note that irreps are labeled as follows. Labels E, T, G, H, etc., refer to irreps that
are, respectively, two-, three-, four-, and five-fold degenerate, etc.  For groups with more than one irrep of a given degeneracy,
subscripts are assigned such that smaller numerical values correspond to more positive character under two-particle exchange. 

From Table~\ref{symtab}, it is clear that as  $N$ increases, an increasingly small fraction of spatial states is allowed by 
permutation antisymmetry. Roughly speaking, this fraction is found to be 
100\%, 84\%,  58\%, 35\%, and 18\%, for $N=2$--6, respectively. In order to exploit this situation as much as
possible, symmetry adaptation of the CCS basis set may be utilized.\cite{klein70,rytter76,herrick83,mcclain,leclerc16} 
 This improves the efficiency of the calculations, by enabling each irrep to be computed independently of the others. 
Additionally, we need only perform calculations for those irreps that are physically realizable.

There are two modes in which symmetry adaptation of the basis can be implemented. The first pertains to 
a general CCS basis of the \eq{CCSbasis} form, for which the individual $X_{m_x}$, $Y_{m_y}$, and $Z_{m_z}$
factors are themselves $S_N$-symmetry adapted.  This is the eventual goal, i.e. use of a CCS basis for which 
the individual component-wise factors are fully correlated across all $N$ particles,  optimized for the specific
application at hand, \emph{and} symmetry-adapted.
More specifically, the single-component, $X_{m_x}(x_1,\ldots,x_N)$, etc., basis functions
will be obtained as the eigenstates of optimal ``effective'' component Hamiltonians, 
$\hat H_x$, etc.,\cite{poirier97b,poirier99qcII,poirier00prec} 
which themselves are invariant under $S_N$.  The effective component Hamiltonian 
problems have one-third the dimensionality 
of the full problem (and can themselves be symmetry-adapted), and are therefore presumed tractable.

Although perhaps a bit counterintuitive at first,
 the resultant ``optimal CCS basis'' (OCCSB) functions  may be conceptualized as the 
\emph{component-based analog of the SCF orbitals}. In other words, whereas the SCF molecular orbital structure 
is what naturally arises when one assumes only separability by particle, the OCCSB is what naturally arises when one 
assumes only separability by Cartesian component. From a practical computational perspective, the chief advantage 
in either case is the same: a greatly reduced separable basis size, in comparison, e.g.,  with a plane wave representation. 

We can estimate the basis size reduction as follows. For an explicit plane-wave calculation, the  basis 
size is $N_B= L^{3N}$.  A plane-wave basis may be a reasonably good choice for certain applications, but for attractive
external Coulomb potentials, it leads to $L \approx 100$ or so, as discussed (Sec.~\ref{CCSHamrep}).
For $N=4$ or more, $N_B$ exceeds one mole in size.  For an OCCSB, the basis size is $N_B=M_d^3$. 
For $N=4$, we estimate that $M_d$ might be in the range 1000--10000, leading to as few as  a ``mere'' 
billion total basis 
functions in all. This is well within reach classically;  CCS calculations to date have been performed with from
$10^{12}$ to $10^{18}$ basis functions\cite{jerke15,jerke18,jerke19,bittner}---although basis size \emph{per se} 
is not the only factor governing overall cost of the calculation.  Further details are provided in an upcoming publication, 
with Sec.~\ref{results} of this paper also providing some new and compelling numerical evidence, in the context of
the harmonium calculations.

In any event, for any CCS basis that respects component-wise $S_3$ symmetry (i.e., whether the OCCSB
or not), symmetry adaptation for the final $3N$-dimensional basis of \eq{CCSbasis}  is straightforward to manage.  In 
effect, each triple-Cartesian product implied by \eq{CCSbasis} has an $S_N$ character decomposition 
that can be determined using two applications of the $S_N$ direct-product table.  In practice then, one simply 
gathers together the triple-product basis functions for a given (physically allowed) irrep, and uses just those basis 
functions to construct a symmetry-adapted block of the Hamiltonian matrix. 

The second mode of symmetry adaptation applies in cases where the $X_{m_x}(x_1,\ldots,x_N)$, etc., functions
are \emph{not} themselves symmetry-adapted. This is necessarily the case if the basis functions are also 
particle-separated, as in \eq{planebasis}.  In such cases, the basis functions naturally partition into equivalence
classes for which the $N$ individual $m_{id}$ indices are identical, apart from permutations. It is then possible
to create symmetry-adapted linear combinations from each such equivalence class, as 
follows:\cite{klein70,rytter76,herrick83,mcclain,leclerc16}   
(a) using the $S_N$ character table, construct the representation of the projection operator for each irrep, in the 
small set of basis functions in a given equivalence class;  (b) apply each projection operator to each product basis 
function in turn, to construct the symmetry-adapted linear combinations. 

The above procedure has been applied for the  $N=3$ and $N=4$ applications presented in Sec.~\ref{results}, which 
represent the first $N>2$ CCS calculations ever performed. Although there are a number of possible
scenarios (reminiscent of quantum statistics) a simple example will serve to clarify the procedure.  Consider
a set of $N=3$ particle-separated basis functions of the \eq{planebasis} form, for which two of the three
indices,  $m_{1x}$, $m_{2x}$, and $m_{3x}$ have the same value $m$, and the third has a different value
$n \ne m$. This defines an equivalence class of three orthogonal basis functions, i.e. 
\ea{
\phi^{x}_{n}(x_1)\phi^{x}_{m}(x_2)\phi^{x}_{m}(x_3) & = & \inprod{x_1x_2x_3}{nmm} \nonumber \\
\phi^{x}_{m}(x_1)\phi^{x}_{n}(x_2)\phi^{x}_{m}(x_3) & = & \inprod{x_1x_2x_3}{mnm} \nonumber \\
\phi^{x}_{m}(x_1)\phi^{x}_{m}(x_2)\phi^{x}_{n}(x_3) & = & \inprod{x_1x_2x_3}{mmn} \nonumber 
}
Construction and application of the irrep projection operators then results in one (unnormalized) linear combination
with A$_1$ character, zero combinations with A$_2$ character, and two combinations forming an E 
pair, as follows:
\ea{
\mbox{A}_1: &  & \ket{nmm} + \ket{mnm} + \ket{mmn} \\
\mbox{E}:  &   &      \ket{nmm} + \ket{mnm} - 2 \ket{mmn}, \\
                   &   &      \ket{nmm} - \ket{mnm} 
}

On reflection, the permutation symmetry adaptation schemes used in the CCS context,
as described above, are certainly much more complicated than the simple and familiar 
SD ``work horse.'' However, they are also much more flexible, as they 
can be applied to any possible combination of spin and spatial states, and moreover,
need not be restricted to overall $A_2$ symmetry character.  This last point is extremely
important, vis-\`a-vis generalizing the CCS method for the combined electron-nuclear
motion quantum many-body problem, wherein electrons and nuclei (be they fermions or
bosons, identical or not) are treated on an equal footing.  Such applications present a
rich, group theoretical structure that cannot be properly addressed simply by using 
SDs.  They are, however, easily amenable to the CCS symmetry adaptation
tools as laid out in this subsection---thereby addressing point (c) from Sec.~\ref{intro}. 

We conclude this subsection with a brief discussion of numerical contamination. On a classical computer,
it is well-known that numerical roundoff error can introduce a tiny contribution 
from the wrong symmetry character into the wavefunction.  Moreover, repeated applications,
e.g. of the Hamiltonian acting on the wavefunction, can magnify this error exponentially.\cite{wang01,montgomery03} 
There is, however, a simple remedy for this problem as well; periodically, the projection operator
for the desired irrep should be applied to the current wavefunction, to ensure its purity.  

Intriguingly, the situation on a quantum computer is quite different. There are, first of all,  various methods
that have been devised for imposing rigorous antisymmetry on qubit representations of the electronic
wavefunction, analogous to the procedures described above.\cite{abrams97,berry18,aspuru20})
Moreover, as in the classical computing case, small symmetry errors are to be expected.   
Interestingly, however, a recent, rigorous error bound characterization\cite{aspuru20} found that
these errors do \emph{not} grow exponentially with repeated applications.  This is a quite important
result, evidently due to the fact that all operations performed on a quantum computer are 
rigorously unitary---unlike the analogous operations on a classical computer.  It is also quite fortuitous,
as the implementation of projection operators that would otherwise be required  to purify the wavefunction
might pose a difficulty on a quantum computer, because projections are necessarily \emph{non}-unitary 
operations. 


\section{Results: Classical Computing Benchmarks}
\label{results}
	
The classical CCS algorithm has  been implemented in the form of our computer code, \emph{Andromeda}, 
and applied to several small benchmark applications, with  $N=2$, 3, and 4 explicit 
electrons..\cite{jerke15,jerke18,jerke19,bittner}   
In every case to date, a plane-wave or sinc basis representation was used.  Results for selected 
earlier and preliminary work are summarized in Sec.~\ref{results-earlier}.
In  Sec.~\ref{results-harmonium}, we present results for the ``harmonium'' system (i.e., with 
harmonic external field $V_{{\rm ext}}$), including earlier $N=2$ calculations,\cite{jerke19} together with
new results for $N=4$.  Other prior or preliminary results, e.g. for long-range 
Ewald potentials,\cite{jerke19} for the H$_2$\cite{jerke18} and H$_3^+$ electronic structure problems, 
and for the H$_2$ and H$_2^+$ combined electron-nuclear  motion problems,\cite{perez13,liu09}
will not be discussed further here.

Collectively,  the above model calculations are useful in their own right, but also in terms of what they 
reveal about prospects for the classical CCS methodology, once optimized OCCSB basis 
sets are employed instead of plane waves.  The model calculations are also quite useful as benchmarks for 
first-quantized QCC calculations---providing more accurate estimates for the required number of 
qubits, for example, than the methods that have been relied upon in the past. 

First, however, we provide a very brief overview of the classical CCS algorithm, as it is currently
implemented in \emph{Andromeda}.  For a much more detailed exposition, please consult
the references.\cite{jerke15,jerke18,jerke19}  To begin with, the CCS tensor-SOP representations for both
the wavefunction vector and the Hamiltonian matrix have already been specified, in \eqs{tensorpsi}{gammasum},
respectively.  Insofar as the algorithm is concerned, 
the main operation is a block Krylov subspace procedure, similar to  
Lanczos.\cite{RN552}  This requires a sequence of matrix--vector products with the Hamiltonian.  
Since both matrix and vector are tensor SOPs, the operation is fast.  On the other hand,
the tensor-SOP rank---or number of terms, $\Lambda$, in the \eq{tensorpsi} sum---increases rapidly with
each successive matrix--vector product.  To curb this growth in $\Lambda$,  the alternating least
squares (ALS) approach of Beylkin and coworkers is used.\cite{beylkin05} 	

\subsection{Summary of previous classical CCS results}
\label{results-earlier}

\underline{H Atom:}\cite{jerke18}  The simplest benchmark calculation possible
is the hydrogen atom ($N=1$).  In (Ref. [\cite{jerke18}]) we computed the ground state
energy of the non-relativistic H atom to an accuracy of $\sim$0.3 millihartree (i.e., much
better than chemical accuracy).  A bare Coulomb potential was used with a plane-wave 
basis, which constitutes a ``worst-case'' choice, as discussed.  A single-coordinate basis
size of $L=81$ was required, corresponding to a second-order full-CI matrix diagonalization
calculation with $M=531,441$ orbitals and $N_B=531,441$ SDs.  

\underline{He Atom:}\cite{jerke18}  In the original study of Ref. [\cite{jerke18}]), an 
explicit $N=2$ calculation of the He atom was also performed, again with bare Coulomb
potentials and a plane-wave basis.  The greater $Z=2$ attraction of the He nucleus as 
compared with H necessitates a somewhat finer grid. Moreover, energy excitation and partial 
screening demand larger coordinate ranges.  Consequently, $L=101$ is now required to 
converge the the  lowest-lying dozen or so electronic states to chemical accuracy.  
This corresponds to a second-order full-CI matrix diagonalization calculation with 
$M=1,030,301$ orbitals and $N_B\approx 1.0615\,\, 10^{12}$ SDs. Note that 
we are able to obtain the excited electronic energies (and wavefunctions) without 
expending any additional numerical effort beyond that required to obtain the ground state.

Finally, in more recent (unpublished) work, 
the plane wave basis was supplemented with additional Gaussians to better handle 
the cusp region.  This enabled reduction of numerical convergence error to tens of 
\emph{micro}hartrees.  

\underline{Li Atom:}  A first-quantized plane-wave calculation of the Li atom ($N=3$)
ground state energy has been touted as an important QCC benchmark.\cite{aspuru20} 
The main reasons are that: (a) the required number of logical qubits has been estimated 
to be at  least 100, and may therefore not be very far beyond present-day limits of quantum hardware, 
and; (b) the classical calculation is  believed to be intractable.  Conversely, if the above Li atom
calculation \emph{is} actually possible on a classical computer, then this finding 
would be quite significant, because it would provide an important first-quantized QCC benchmark, while simultaneously 
pushing back the threshold for ``quantum supremacy.''  

Using our classical CCS code \emph{Andromeda}, we have obtained preliminary results 
for the Li atom---again, treating all electrons explicitly, and using bare Coulomb potentials and  a
plane-wave basis.  Our preliminary results suggest  not only that a classical calculation 
is possible, but also that a chemically accurate QCC calculation (performed using exact 
$\gamma$ matrices; see Sec.~\ref{quantumCCS})  should require only $\sim$60 logical qubits
($3Nn=63$, with $N=3$ and $n=7$).
This places the Li atom application within much closer reach of 
present-day quantum hardware than might have been imagined.  
On the other hand, our assertion is not yet fully confirmed, as it is based on an 
$L=101$ ($M=1,030,301$, $N_B = 1.0937\,\,10^{18}$) classical calculation that
is not fully completed.

The reason is because of very slow convergence with respect to the number of 
Krylov subspace iterations. More specifically, we have not yet been able to achieve a 
convergence of the  Li atom ground
state energy to better than a few tens of millihartree---it seems that a linear space of  
$N_B = 10^{18}$ dimensions is simply too large  for such methods to operate 
effectively within.  Note that memory limitations are decidedly \emph{not} the issue here!
Eqs.~(\ref{tensorpsi}) and (\ref{gammasum}) have no difficulty reducing RAM
requirements for such astronomical $N_B$ values down to a very modest size. 

In any event, although the computed ground state 
energy is not yet Krylov-converged, there is an independent indication
 that the basis set convergence itself is sufficient to achieve at least 
near-chemical accuracy.  More specifically, \emph{Andromeda} has the 
capability to compute the total probability of a computed wavefunction along each of
the $6N$ position coordinate edges (i.e., minimum and maximum $d_i$ grid values), 
\emph{and also} each of the $6N$ momentum space edges. According to this
metric, the largest  position-edge probability for the ground state wavefunction
is  $5.60\,\, 10^{-5}$, whereas the largest momentum-edge probability is 
$2.36\,\, 10^{-8}$.  These values are on a par with similar basis-set converged calculations
for He and H as described above. 

\underline{Correlated Electron Gas (CEG):}\cite{bittner}  The above examples clearly demonstrate that plane 
waves are not well suited  for systems with attractive bare Coulomb potentials.  Even so, we have been able to
perform explicit all-electron calculations of this kind for systems with up to $N=3$ and $N_B = 10^{18}$. 
How much further could we scale up in $N$, if a more ``competitive'' basis such as OCCSB were employed?
Short of actually performing such calculations, the next best thing would be to find systems for which the
plane-wave basis itself might be nearly optimal.   
For this purpose, the best application is probably the ``correlated electron gas'' system---i.e., 
with $V_{
{\rm ext}}=0$, periodic boundary conditions, and a fixed number of electrons per unit cell
(in contrast to the  uniform electron gas or ``jellium'' model\cite{ceperley80,esteban12,varas16}).

Using Ewald summations, we are  computing exchange-correlation energies for the correlated
electron gas system with $N=2$, 3, and 4 explicit
electrons, for both spin-restricted and spin-unrestricted cases.  An extremely broad range of Wigner-Seitz 
radii is being considered, ranging from metallic to Wigner crystal regimes where correlation effects dominate. 
This project is still ongoing.\cite{bittner} However, the best indication is that $L=11$ suffices for most
Wigner-Seitz radius values.  
As anticipated, this represents a  \emph{very} dramatic reduction in basis size---i.e. down to $M=1331$ or 
$N_B = 3.138\,\, 10^{12}$ (for $N=4$).   Put another way, such small $L$ values suggest that explicit
classical CCS calculations up to $N=6$ may be possible. 

\subsection{Harmonium}
\label{results-harmonium}

The Hamiltonian for the harmonium system is identical to that of the standard electronic structure
form of \eq{Ham}, except that the attractive bare Coulomb external potential is replaced with the 
following confining harmonic field:\cite{RN756,RN773,RN755,jerke19}
\eb
	V_{{\rm ext}}(x_i,y_i,z_i)  = {\omega^2 \over 2} \of{x_i^2 + y_i^2 + z_i^2} \label{harmonium}
\ee
As a technical matter, too, we consider harmonium with a Yukawa, rather than Coulomb, form
for the pair repulsion potential:
\eb
V_{ee}(\vec r_i ,\vec r_j )  = e^{-\gamma |\vec r_i-\vec r_j|} /  |\vec r_i-\vec r_j| \label{yukawa}
\ee

One of the remarkable features of harmonium---or the ``Hooke atom,'' as it is also called---is that 
(nearly) analytical solutions exist for the $N=2$ case, 
despite the coupling implied by the pair repulsion contribution. 
This makes it possible to compare numerically computed results  directly with  ``exact'' values, 
thus offering an excellent opportunity to benchmark new methods.  

For our purposes, 
we are also interested in harmonium as an ``intermediate'' test case, lying in between the 
pathological Coulomb $V_{{\rm ext}}$ form of the atomic models, and the $V_{{\rm ext}}=0$ form
of the correlated electron gas model (Sec.~\ref{results-earlier}).   In particular, we expect that the
plane-wave basis---though not an especially good choice for harmonium---will nevertheless perform
far better here  than for bare Coulomb potentials. Of course, there are many contexts
in electronic structure in which the bare Coulomb interaction is replaced with ``milder'' 
forms---e.g., screened potentials (including SCF), effective core potentials, ``softened'' Coulomb
potentials, etc.    

In Ref.[\cite{jerke19}] we used \emph{Andromeda} to perform a classical CCS calculation of
the 20 lowest-lying  energy eigenstates (i.e. energy levels and wavefunctions) for the $N=2$
harmonium system with $\omega=1/2$  and  $\gamma=1$ (in atomic units).  This parameter choice    
guarantees ``non-trivial''  behavior---with neither the external confining field nor the interparticle repulsion
contribution dominating.  By comparing with the exact analytical results, we were able 
to confirm an overall numerical convergence of all states to better than a few tens of
\emph{micro}hartrees.   Moreover, this was achieved using a \emph{much} smaller basis 
than in the atomic examples of Sec.~\ref{results-earlier}---i.e., $L=33$ (or $M=35,937$,
$N_B=1.2915 \,\, 10^{12}$).  For chemical accuracy, the $L$ value could be reduced by 
about another  factor of two.   

For the present study, we have extended the above harmonium calculations out to $N=4$,
using the same $\omega$ and $\gamma$ values.  
More specifically, we have computed the lowest-lying 15 states belonging to the A$_1$ 
and T$_2$ irreps, with
the goal of achieving chemical accuracy.  The results, together with Krylov convergence
errors, are presented in Table~\ref{harmoniumtab}.  By this measure, near chemical accuracy is 
indeed achieved (rms error = 1.8 millihartree), 
using  a single-coordinate basis size of only $L=17$ (i.e., as 
predicted by the $N=2$ calculation). This corresponds to $M=4913$ or  $N_B\approx 5.8262 \,\,10^{14}$.

For the $N=4$ harmonium ground state energy, the Krylov convergence 
error is only $\sim$0.4 millihartree.  For this
state, we have also used \emph{Andromeda} to compute edge probabilities (as  
for the Li atom in Sec.~\ref{results-earlier} above).
The largest  position-edge 
probability is found to be only $2.43 \,\, 10^{-7}$; the largest momentum-edge probability is only 
$1.74\,\, 10^{-9}$.  These values imply a very small basis set truncation error---which, in any 
event, is  likely to be much smaller than other sources of numerical error for this 
calculation.

\begin{table}[h]
\small
  \caption{Fifteen lowest-lying energies for the $N=4$ harmonium system (atomic
  units), as computed using the \emph{Andromeda} classical CCS code, for the A$_1$ and
  T$_2$ irreps (Column III).  Numerical convergence 
  errors (Column IV) are taken relative to the previous Krylov iteration. $S_4$ irrep labels (Column II) refer
  to the spatial part of the wavefunction only.}    
  \label{harmoniumtab}
  \begin{tabular*}{0.48\textwidth}{ccccc}
    \hline
    State index & Spatial irrep  &  Energy & Error \\
    \hline
    1 & A$_1$  &   3.6341  &   -0.0004  \\
    2 & A$_1$  &   3.9406  &  -0.0010 \\
    3 & A$_1$  &   3.9572  &  -0.0025  \\
    4 &  A$_1$  &   3.9663  &  -0.0017 \\
    5 & T$_2$   &   4.1045  &  -0.0007 \\
    6 & T$_2$   &   4.1288  &  -0.0010 \\ 
    7 & T$_2$   &   4.1349  &  -0.0021 \\
    8 & A$_1$   &  4.3455   &  -0.0029  \\
    9 & A$_1$   &   4.3668   &  -0.0048 \\
    10 &  A$_1$&   4.4304  &  -0.0025 \\
    11 &  A$_1$ &   4.4540   & -0.0036  \\
    12 &  A$_1$ &   4.4799  &  -0.0036 \\
    13 &  T$_2$ &   4.4945  & -0.0016  \\
    14 & T$_2$  &    4.4999  &  -0.0020 \\
    15 &  T$_2$ &   4.5057   &  -0.0021 \\
    \hline
  \end{tabular*}
\end{table}


\section{Results: Quantum Computing Implementation}
\label{quantum}

As discussed, in ``first quantized'' QCC,\cite{abrams97,zalka98,lidar99,abrams99,nielsen,aspuru05,kassal08,whitfield11,huh15,babbush15,babbush17,kivlichan17,babbush19,aspuru20} 
the wavefunction is represented explicitly, 
as a function over the entire $3N$-dimensional configuration space of electron positions. 
A set of logical qubits is used for this purpose---e.g., to represent the value of the wavefunction
at each of a set of uniformly-spaced sinc grid points, forming a $3N$-dimensional lattice.
An initial wavefunction is generated, and 
then manipulated through specialized quantum circuitry, in order to simulate the action
of the Hamiltonian.  Eventually, through many such manipulations, the time-dependent
evolution of the initial wavefunction is simulated---or alternatively, the energy eigenstates are
computed. 

For the types of quantum chemistry applications for which QCC methods can currently
be implemented, using today's quantum (or hybrid quantum/classical) computers, 
the second-quantized methods  have the upper hand.  
This is because such calculations use the standard
simplifying approximations, wherein the number of orbitals, $M$, and/or or the number
of SDs, $\of{M \atop N}$, is greatly  reduced from the CBS-limit and/or full-CI  values.  This yields a greatly
simplified calculation requiring exponentially less computational effort than a fully 
converged calculation---which is generally not currently feasible on either classical or 
quantum computers except for very small systems.  As a consequence, the first-quantized 
methods require more logical qubits, to an extent that is still a bit
out of reach, even for simple applications.

On the other hand, the list of such simple  
applications that have actually been realized using second-quantized  QCC---which includes, 
e.g. crude ground state calculations for the H$_2$ molecule and other two-electron 
systems\cite{izmaylov19,aspuru20}---still remains short and unimpressive in comparison
with state-of-the-art quantum chemistry on classical computers.\cite{shepard14,shepard19,lischka20} 
In the coming years, as quantum computing technology develops and larger QCC
applications can be realized, we will reach the point where first-quantized methods 
overtake second-quantized methods, vis-\`a-vis their efficiency and accuracy on 
quantum computers, and their ability to be run solely on \emph{bona fide} quantum computers
as opposed to hybrid platforms. 

 As discussed, the  primary feature here is that the cost to store the wavefunction, 
in terms of the required number of logical qubits, scales only \emph{linearly} with $N$.  
Our first task in this section, then, is to estimate the number of logical qubits required. 
Our second task shall be to explain the quantum circuitry used to implement the 
above-referenced Hamiltonian-based manipulations of the wavefunction, in precise detail.  
This we shall do both for the ``standard'' or canonical Trotter approach,
and also for our alternate CCS-based Trotter implementation, to which the former
will be compared. 

\subsection{Explicit Quantum Representation of the Wavefunction}
\label{CCSquantumwaverep}

Note that both tasks above depend intimately on the choice of basis set. For convenience
and simplicity, we shall throughout this section presume either a  plane-wave, or a sinc 
(plane-wave dual), basis representation---with the latter related to the former via an inverse Fourier
transform. Note that both basis representations conform to \eq{planebasis}.   
Let $\ket{\psi}$ denote the $3N$-dimensional numerical wavefunction vector, whose 
individual components are denoted
\eb
	\Psi_{m_x m_y m_z} = \Psi_{m_{1x},\ldots,m_{Nx},m_{1y},\ldots,m_{Ny},m_{1z},\ldots,m_{Nz}}
\ee
from \eqs{planebasis}{directpsi}.
There are clearly $L^{3N}$ such components in all.  

The total number of logical qubits needed to 
represent the entire $\ket{\psi}$ is thus $3Nn$, with $L=2^n$. Thus, for $L \approx 100$, we can
expect $n \approx 7$.  Note that for each of the $3N$ coordinates, $d_i$, there is a specific 
set of $n$ qubits, used to represent the $L$ corresponding one-dimensional basis 
functions, $\phi^{d}_{m_{id}}(x_i)$.  We therefore associate each such bundle
of $n$ qubits with the coordinate $d_i$, but, also with the part of the wavefunction associated 
with that coordinate, which we call $\ket{\psi_{id}}$. Note that the concept of $\ket{\psi_{id}}$
has meaning in quantum computing, \emph{even when} $\ket{\psi}$ itself is not separable \emph{per se}.  

In quantum computing circuit diagrams, individual qubits are traditionally represented as horizontal ``wires''.
These flow from left to right, encountering various quantum gates along the way that alter the qubit states---and
thereby, also, the corresponding  $\ket{\psi}$.  In addition to the logical qubits representing  $\ket{\psi}$, various
``ancilla'' or workspace qubits are also often relied upon, to be used at various stages throughout the
quantum computation. These are often used to get around unitarity restrictions---although here, the ancilla
qubits generally represent function outputs.  Ancilla qubits are typically presumed to be in an initial 
(and often final) state of ``zero'', which we denote as $\ket{0}$.  

Some additional quantum circuit conventions that we adopt here are described next. 
Since for each dimensional set, the number of qubits, $n$, is fixed---and since there is no reason to consider 
subdividing below the level of a single coordinate---in all of the quantum circuit diagrams presented here, 
we treat connections involving $\ket{\psi_{id}}$ as if they were \emph{single-qubit wires}, rather
than bundles.
As additional notation, 
let $\ket{\psi_i}$ represent the  part of $\ket{\psi}$
associated with particle $i$, spanning all three Cartesian coordinates, $x_i$, $y_i$, and $z_i$.
Likewise, $\ket{\psi_{ij}}$ includes all Cartesian coordinates for particles $i$ and $j$, whereas
$\ket{\psi_{ijd}}$ includes just the two $d$'th components---e.g., $\ket{\psi_{12x}}$,  
associated with  $x_1$ and $x_2$.  Also, whereas $\ket{\psi}$ refers to the wavefunction in the 
sinc representation, the corresponding Fourier or plane-wave representation is denoted $\ket{\tilde \psi}$.
Finally, the usual ``slash'' notation will be used for ancilla qubit bundles,  representing function 
outputs.  To achieve chemical accuracy, each such bundle will typically require around
15--20 qubits.  

\subsection{Quantum Circuitry: Canonical Trotter Method}
\label{quantumTrotter}

\begin{figure*}[htb]
 \centering
 \includegraphics[height=9cm]{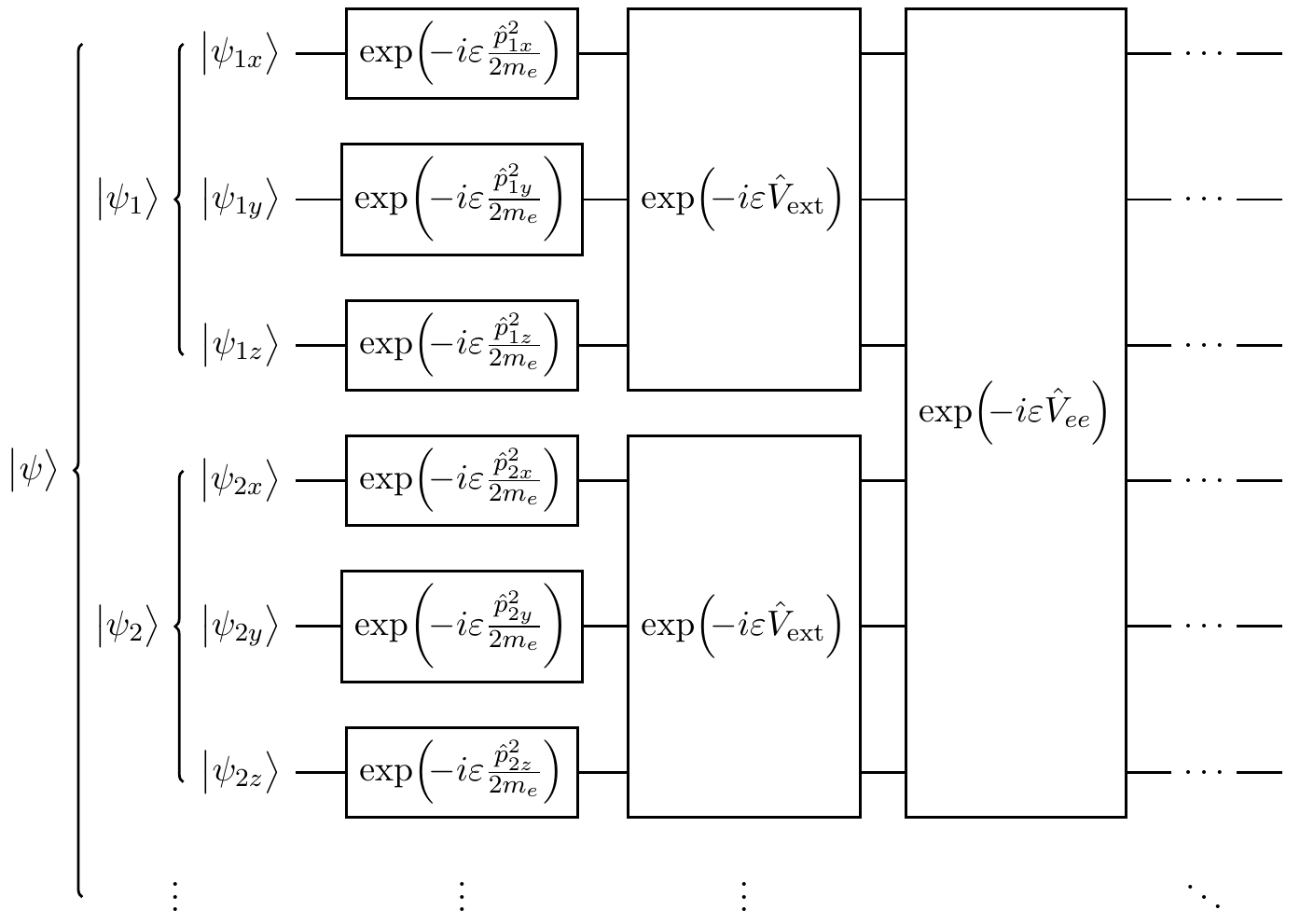}
 \vskip-5pt
 \caption{Overview of quantum circuit to implement  unitary short time evolution, $\hat U(\ep)\ket{\psi}$, via Trotterization, on a quantum computer.  This structure is used in both the canonical and CCS Trotter quantum circuitry.}
 \label{fig1}
\end{figure*}

Next we address the second task above, i.e. describing the quantum circuitry necessary to 
implement the requisite Hamiltonian manipulations of $\ket{\psi}$.
For both time-independent and time-dependent QCC applications, one must apply 
unitary time evolution of the form $\ket{\psi(t)} =\hat{U}(t)  \ket{\psi(0)}$,\cite{aspuru05,nielsen,aspuru20} 
where $\ket{\psi(0)}$ is the initial wavefunction.
In the time-dependent context, this directly propagates the wavefunction, as desired.  In the 
time-independent context, QPE\cite{abrams99,nielsen,parrish19,izmaylov19,aspuru20}
 is used to obtain the ground eigenstate (and in some cases, 
excited eigenstates); however, this too requires unitary time evolution of the above 
form.\cite{abrams99,nielsen,parrish19,izmaylov19,aspuru20}

The specific form of the time-evolution operator is $\hat{U}(t) = \exp{\sof{-(i/\hbar) \hat H t}}$, 
where $t$ is the final time, and $\hbar$ is the reduced Planck's constant (taken to be unity in all that follows).
In the canonical Trotter method,  this  is implemented via repeated applications of $\hat U(\ep)$, where 
$\ep$ is a sufficiently small time step such that the  individual contributions from 
\eq{Ham} can be treated separately. In other words, 
\ea{
     \hat U (\ep) & \approx &  
     \cdots  \exp\!\sof{\!-i\ep {\hat V}_{ee} (x_1,y_1,z_1,x_2,y_2,z_2)} \times    \label{Trotter} \\
     & & \cdots  \exp\!\sof{\!-i\ep {\hat V}_{{\rm ext}} (x_1,y_1,z_1)}          \times  \nonumber  \\
    & &  \cdots \exp\!\sof{\!-i\ep {{\hat p}^2_{1z} \over 2 m_e}} \exp\!\sof{\!-i\ep {{\hat p}^2_{1y} \over 2 m_e}} 
    \exp\!\sof{\!-i\ep {{\hat p}^2_{1x} \over 2 m_e}}, \nonumber
}
an approximation known as ``Trotterization,'' or the Lie-Trotter-Suzuki 
decomposition.\cite{trotter59,zalka98,nielsen,georgescu14,kivlichan19,aspuru20}
An overview of the quantum circuit implementation of $\hat U(\ep)\ket{\psi}$, 
based on \eq{Trotter}, is presented in Fig.~\ref{fig1}.   Note that the associated ancilla bits,
needed to carry out the individual operations in practice, are not indicated in this figure.  

We next turn to the canonical Trotter implementation of the various Hamiltonian component 
contributions in \eq{Trotter} and Fig.~\ref{fig1}.  To begin, we note that it is easiest to implement
matrix--vector products on a quantum computer, if the matrix is diagonal.  Then, the additional
condition of unitarity implies that each $\ket{\psi}$ component is simply multiplied by some
phase shift.  However, we further note that, due to basis set truncation error, the sinc basis representation is only 
an \emph{approximate} position representation. Thus, in reality,  an exact sinc-function representation
of the potential energy contributions in \eq{Trotter} gives rise to matrices that are only \emph{nearly} 
diagonal.  More specifically, we find that  the sinc functions themselves are not true Dirac delta
functions, but rather, narrowly peaked functions centered around a uniformly-distributed set of 
discrete grid points, known as the ``sinc discrete variable representation'' (sinc DVR) grid 
points.\cite{RN553,colbert92,szalay96,light00,poirier02dvrlj,littlejohn02b}
 
 We are therefore led to a natural, diagonal, grid-based \emph{approximation} to the true potential matrices,
 obtained by simply evaluating the potential function
 at the individual sinc DVR grid points, taking these values to comprise the diagonal matrix elements,
 and then setting the off-diagonal elements to zero.\cite{poirier02dvrlj}  Note that this procedure introduces
 new ``quadrature error,'' above and beyond the basis set truncation error already 
 present.\cite{RN553,colbert92,szalay96,light00,poirier02dvrlj,littlejohn02b}  Coulomb 
 potentials---being both singular and long-range---present a ``worst-case scenario'' in terms of sinc DVR 
 quadrature error.  For example, one must avoid placing sinc DVR grid points directly on Coulombic singularities, or
the sinc DVR matrix elements become infinite!  On the other hand, the basis set truncation errors associated 
with sinc or plane-wave representations of Coulomb potentials also present a worst-case scenario, 
as discussed previously. 

In any event, it is in this fashion that the potential energy contributions in \eq{Trotter} are implemented---i.e., 
as phase shifts obtained from potential energy evaluation at the sinc DVR grid points.  Put another way,
since the potential energy matrices are diagonal in the sinc DVR, their contribution
to the time evolution  can be implemented using ordinary function evaluation---a standard operation 
for quantum computers.\cite{abrams97,zalka98,nielsen,kais,preskill18,alexeev19}
Note that each such 
function evaluation involves only a subset of coordinates---and therefore only a subset of qubits, as indicated
in Fig.~\ref{fig1}. Specifically, each external ($V_{{\rm ext}}$) or pair repulsion ($V_{ee}$) potential energy term 
involves three or six coordinates, respectively.   
 
\begin{figure}[h]
\centering
  \includegraphics[height=4.5cm]{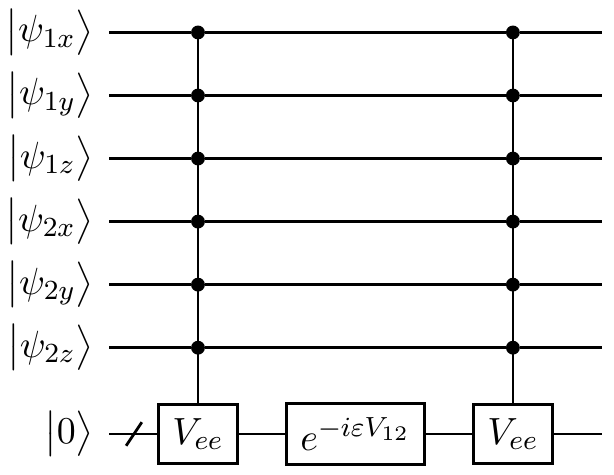}
  \vskip-5pt
  \caption{Canonical quantum circuit to implement $\exp\!\of{\!-i\ep {\hat V_{ee}}}\ket{\psi_{12}}$ 
on a quantum computer. The function evaluation, $V_{12}=V_{ee}(x_1,y_1,z_1,x_2,y_2,z_2)$, involves six explicit coordinates
and sets of qubits, and includes an inverse square-root evaluation.}
  \label{fig2}
\end{figure}

Based on the above description in terms of sinc DVRs, the quantum circuit  used to implement  
$\exp{(-i \ep \hat V_{ee})}\ket{\psi_{12}}$  is given in Fig.~\ref{fig2}. A bundle of ancilla qubits (lowest wire) 
is used to represent the value of the function, $V_{12}=V_{ee}(x_1,y_1,z_1,x_2,y_2,z_2)$, which becomes the 
output of the first gate encountered---i.e., the $V_{ee}$ gate indicated near the lower-left corner. Note that
this gate also receives input from six explicit coordinates or sets of qubits---i.e., $6n$ qubits in all, in addition 
to the ancilla bundle.  These act as ``control qubits,'' associated with specific sinc DVR grid points, whose 
purpose is to determine the associated function output value.  Once
the latter value is determined, it is then directed to the second gate, where it is used to effect a phase shift. 
Note that  as a matter of notation, we use $e^{-i \ep V_{12}}$ to refer to the phase shift gate in the quantum circuit
diagram, whereas $\exp{(-i \ep \hat V_{ee})}$ refers to the corresponding quantum operator.  
Finally, a second $V_{ee}$  function evaluation is applied, in order to ``reset'' the ancilla qubits.   

Implementation of the  $V_{ee}$ function evaluation gate is complex.  Not only does 
this gate receive input from many control and ancilla qubits, but the 
standard pair repulsion function itself is also highly correlated, and involves Coulombic 
singularities that must be carefully avoided.  In addition to multiplications and additions,
which are fairly straightforward to implement (even error-corrected) 
on a quantum computer,\cite{draper00,florio04,gidney19}
the $V_{ee}$  function  also requires evaluation of an inverse square-root, which at present is extremely 
costly if high accuracy is required.\cite{babbushcomm}  
All of this implies a large circuit complexity---rendering the $\exp{(-i \ep \hat V_{ee})}$ 
operation the computational bottleneck of the entire canonical Trotter approach 
(as indeed, is also generally  the case for classical algorithms).  

For $\exp{(-i \ep \hat V_{{\rm ext}})}$, a similar, but simpler circuit than Fig.~\ref{fig2} may be used, for which there are only 
three explicit coordinates and thus three sets of control qubits, associated with a single particle. 
However, this still represents greater complexity than is desired.  As a design goal, 
we seek to reduce the complexity of the calculation down to a few operations involving  no more than one or two qubit sets:
\begin{enumerate}
\item  operations that involve a \emph{change of representation} (e.g., QFT) should
be efficiently reducible to a \emph{single} qubit set.  
\item \vskip-5pt operations that do \emph{not} involve a change of representation should be  efficiently reducible to 
\emph{two} qubit sets.
\end{enumerate}

\begin{figure}[h]
\centering
  \includegraphics[height=2.0cm]{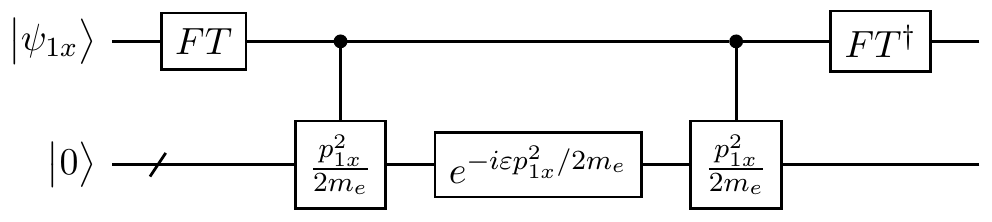}
  \vskip-5pt
  \caption{Quantum circuit to implement $\exp\!\sof{\!-i\ep {{\hat p}^2_{1x} \over 2 m_e}} \ket{\psi_{1x}}$
on a quantum computer. The function evaluation involves just a single coordinate (the momentum $p_{1x}$),
and hence just a single qubit set. This circuit is used in both the canonical and CCS Trotter quantum circuitry.}
  \label{fig3}
\end{figure}

It remains  to consider the kinetic energy contributions to \eq{Trotter}. From Fig.~\ref{fig1}, we observe 
that each such  contribution  satisfies Condition 1. above; only a single qubit set is involved. This is, of course,
because of the fact that the kinetic energy is separable by both particle and Cartesian component.  
On the other hand, the ${\hat p}^2_{id}/(2m_e)$ contributions are decidedly \emph{not} diagonal in the
sinc representation. A Fourier transform is thus required---to transform from the sinc
 to the plane wave representations, in terms of which the matrix representation of 
${\hat p}^2_{id}/(2m_e)$ becomes diagonal.  This is performed using the  standard
quantum Fourier transform (QFT) algorithm,\cite{shor94,abrams99,nielsen} denoted ``$FT$'' in the quantum
circuit diagrams.  Specifically, $FT$ is applied to $\ket{\psi_{id}}$, which thus becomes  $\ket{\tilde \psi_{id}}$.
Then, the $e^{-i \ep p^2_{id}/2m_e}$ phase shift can be implemented, in similar fashion to  Fig.~\ref{fig2}.
The quantum circuit  implementation for this entire $\exp\!\sof{\!-i\ep {{\hat p}^2_{1x} \over 2 m_e}} \ket{\psi_{1x}}$
operation is indicated  in Fig.~\ref{fig3}, for $i=1$ and $d=x$. 
 
\subsection{Quantum Circuitry: CCS Trotter Method}
\label{quantumCCS}

For the CCS implementation proposed here, much of the canonical Trotter quantum circuitry
can be retained. In particular, from a high-level perspective, the overview as presented in 
Fig.~\ref{fig1} applies equally well to both the canonical and CCS Trotter quantum strategies.   Our
CCS implementation as presented here is thus also a Trotterized approach.  Furthermore, 
action of the kinetic energy operator is still implemented exactly as in Fig.~\ref{fig3}, as it
adheres to our ``design philosophy'' as described in Sec.~\ref{quantumTrotter}.

On the other hand, the CCS implementations for $\exp{(-i \ep \hat V_{{\rm ext}})}$, and 
for the bottleneck $\exp{(-i \ep \hat V_{ee})}$ operation, are very different from the canonical
approach. Specifically, we pursue a quantum circuit implementation in keeping with the CCS 
tensor-SOP decomposition described in Sec.~\ref{CCSHamrep}. 
For brevity, we focus the discussion that follows on the more challenging case of implementing
$\exp{(-i \ep \hat V_{ee})}$, although similar procedures can also be applied for 
$\exp{(-i \ep \hat V_{{\rm ext}})}$.  

\begin{figure}[h]
\centering
  \includegraphics[height=5.5cm]{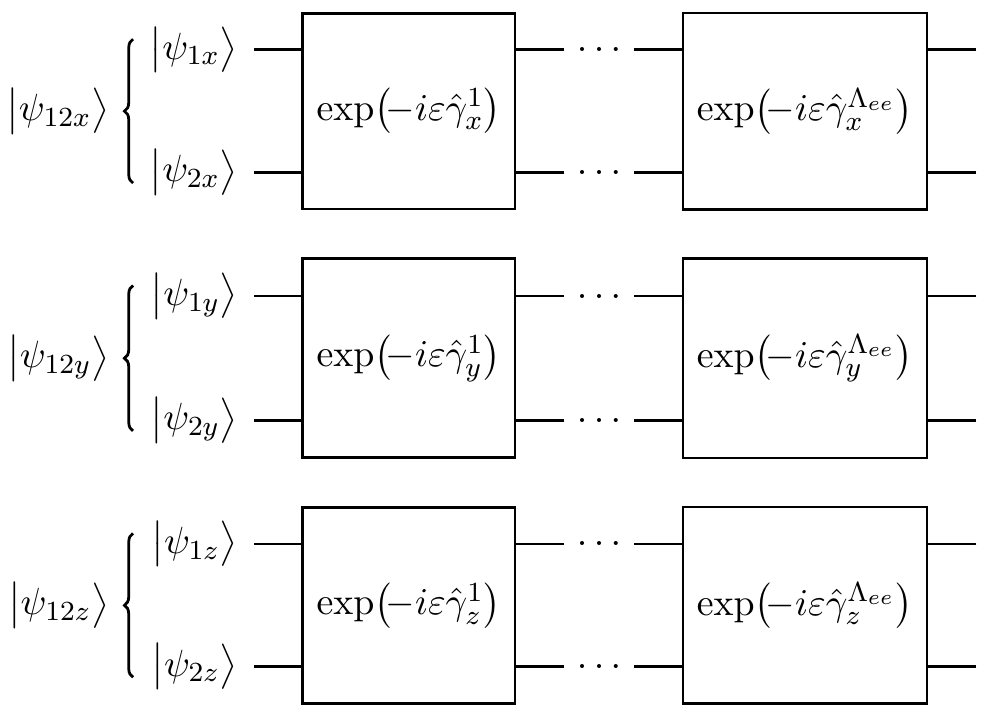}
    \vskip-5pt
  \caption{CCS quantum circuit to implement $\exp\!\of{\!-i\ep {\hat V_{ee}}}\ket{\psi_{12}}$ 
on a quantum computer, based on \eq{gammasum}. The six-dimensional $V_{ee}$ function gate of Fig.~\ref{fig2}  
is now replaced with two-dimensional  function gates, which separate by Cartesian component. 
 The two-particle  wavefunction  contribution 
$\ket{\psi_{12}}$ decomposes naturally into  $\ket{\psi_{12x}}$, $\ket{\psi_{12y}}$, and $\ket{\psi_{12z}}$
components.}
  \label{fig4}
\end{figure}

From the CCS tensor-SOP form of \eq{gammasum}, it is clear that a natural
quantum circuit implementation for  
$\exp\!\of{\!-i\ep {\hat V_{ee}}}\ket{\psi_{12}}$ should take the form presented in Fig.~\ref{fig4}. 
In comparing with the canonical quantum circuit of Fig.~\ref{fig2}, we see that the two 
implementations are completely different.  The CCS version would  appear to  
offer many advantages. In particular, the six-dimensional function calls are
replaced with $\exp\!\of{\!-i\ep {\hat \gamma}^\lambda_d} \ket{\psi_{12d}}$ operations involving 
just \emph{two} coordinates (e.g. $x_{1}$ and $x_{2}$)---thereby satisfying design Condition 2. above. 
Moreover, the $\gamma_d^\lambda$ functions themselves are  Gaussians, with very smooth behavior 
and a restricted, well-defined range.  

\begin{figure*}[htb]
 \centering
 \includegraphics[height=2cm]{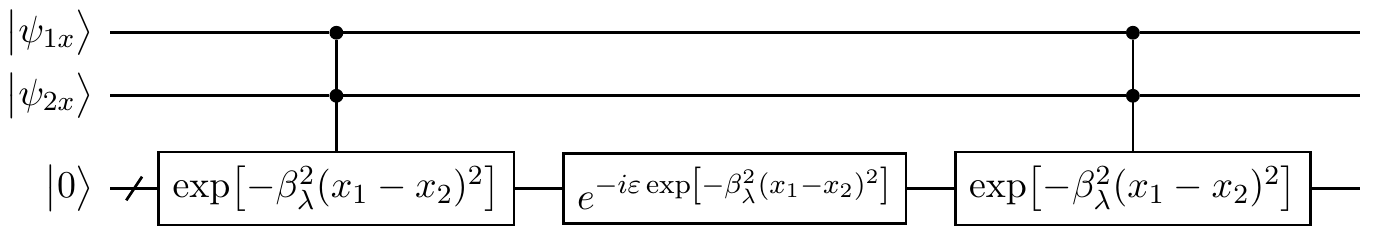}
 \vskip-5pt
 \caption{CCS quantum circuit to implement $\exp\!\of{\!-i\ep {\hat \gamma^\lambda_{x}}}\ket{\psi_{12x}}$ 
on a quantum computer, in the sinc DVR. The function evaluation, 
$\gamma_x^\lambda = \exp\!{\sof{-\beta_\lambda^2 (x_1-x_2)^2}}$, 
involves two explicit coordinates and sets of qubits, and includes an exponential function evaluation.}
 \label{fig5}
\end{figure*}

As implied above, the ${\hat \gamma}^\lambda_d$ matrix representations are  diagonal in the sinc DVR,
and thus amenable to a straightforward quantum circuit implementation---i.e., essentially 
a two-dimensional version of Fig.~\ref{fig2}. The specific quantum circuit used to 
implement  $\exp\!\of{\!-i\ep {\hat \gamma^\lambda_{d}}}\ket{\psi_{12d}}$ in the sinc DVR is presented in Fig.~\ref{fig5}. 
Note that a similar quantum circuit may be employed for the CCS tensor-SOP components of
$V_{{\rm ext}}$, except that only a single coordinate or set of control qubits is needed. 

Diagonal sinc DVR ${\hat \gamma}^\lambda_d$ matrices
certainly present one reasonable avenue for QCC implementation of the CCS approach. 
Their Gaussian form, moreover, implies that far fewer grid points would be  needed to represent $ \hat V_{ee}$ 
 than in the canonical approach (to the same  level of accuracy).  On the other hand, a  sinc DVR 
 calculation still does introduce some quadrature error, which does not exist in the \emph{classical} CCS 
implementation---e.g., as was used to generate the results of Sec.~\ref{results}.   For this reason,
it might be less efficient, in terms of basis size, than the corresponding classical calculation.     
That said, insofar as factors determining the necessary basis size  is concerned, 
$\hat V_{{\rm ext}}$ is \emph{much} more important than $ \hat V_{ee}$, if bare Coulomb potentials are used. 
Consequently, it is not so clear how much worse this CCS form of quadrature error would actually make 
things, in practice.

Nevertheless, we maintain that it is worth the effort to try to develop alternate QCC implementations 
that use \emph{exact} sinc ${\hat \gamma}^\lambda_d$ matrices (i.e., as computed 
without recourse to sinc DVR grid points)---thereby eliminating all quadrature error, and reducing
the basis size still further,  down to exactly the number used in classical calculations.  In this context, the 
greatest challenge is posed by the fact that the exact matrices are no longer diagonal.  Also, though  
still quite small by virtue of being two-dimensional, they are not quite small enough to satisfy
our  design conditions.    
 
Despite the above challenges, we have developed an exact sinc QCC implementation that
not only overcomes the non-diagonal limitation, but also satisfies both design conditions.  
 This approach is based on the fact that the off-diagonal part of an exact $\hat \gamma$ matrix 
 can be written in the form
${\hat \gamma}_{{\rm offdiag}}=( \hat {p_i} \hat{p_j} \hat{D} +  \hat{D} \hat {p_i} \hat{p_j})/2$---where 
$\hat D$ is diagonal,  and $d$ subscripts (and $\lambda$ superscripts) have been dropped for clarity. 
In the $\ep \rightarrow 0$ limit,  Trotterization then yields
\eb
\exp\!\of{\!-i\ep {\hat \gamma}_{{\rm offdiag}}} \approx  
     \exp\!\of{\!-i\ep  \hat {p_{i}} \hat{p_j} \hat{D}/2}  
     \exp\!\of{\!-i\ep  \hat{D} \hat {p_i} \hat{p_j} /2}.
     \label{nonunitary}
\ee

Note that the  individual factors in \eq{nonunitary} above are  \emph{no longer unitary}; the 
exponents are comprised of non-commutative \emph{operator}---rather than tensor---products.
Although some algorithms have come on the scene very recently,\cite{rasmussen20,endo20}
we have developed our own QCC implementation that is unitary, and correct to first
order in $\ep$. It involves separate $FT$ operations for $\hat {p_1}$ and $\hat {p_2}$,
and also a two-dimensional function evaluation for $\hat D$---thereby satisfying our two design conditions. 
As this work is currently under review by Texas Tech University for possible patent protection, 
further details will be presented in a future publication.


\section{Conclusions}
\label{conclusions}

The frontier in quantum chemistry calculations has always involved pushing 
the following limits: (a) full CI; (b) complete basis set (CBS); (c) number of  explicit electrons, $N$.
All of these imply a substantial increase in $N_B$, the basis size or number of Slater Determinants 
(SDs) needed to perform the calculation.  Over the decades---but especially in the last 10 years or 
so---impressive and steady progress has been 
made towards increasing the number of SDs that can be treated explicitly in calculations.\cite{RN544,hohenstein12,parrish13,shepard14,shepard19,lischka20,muller09,jong10,vogiatzis17}
Much of the recent progress has relied on massive parallelization, but use of tensor decompositions of 
various kinds is also playing an increasingly important role.   Moreover, going forward, it is clear that
quantum computing---for which exponential growth in $N_B$ corresponds
to mere \emph{linear} growth in the number of logical qubits---will engender an even greater 
sea change.  This long-awaited ``quantum computational chemistry (QCC) revolution''\cite{poplavskii75,feynman82,lloyd96,abrams97,zalka98,lidar99,abrams99,nielsen,aspuru05,kassal08,whitfield11,brown10,christiansen12,georgescu14,kais,huh15,babbush15,kivlichan17,babbush17,babbush18,babbush18b,babbush19,low19,kivlichan19,izmaylov19,parrish19,altman19,cao19,alexeev19,bauer20,aspuru20} may be nearly upon us,  although 
achieving full quantum supremacy will likely  have to wait for quantum platforms that can accommodate first-quantized 
methods. 

Whatever the future may bring, it seems  that the quantum chemistry discipline may be at a ``tipping point''---a nexus, 
where there is once again room to consider  bold new ideas.  One such idea---that the electronic structure
Hamiltonian operator separates more naturally by \emph{Cartesian component} than by particle---represents a 
radical departure from the traditional particle-separated starting point, and serves as the focus of this work.  
In addition to directly addressing the traditional three limits, (a)--(c) above, the CCS approach also has
ramifications for two additional quantum chemistry frontiers: (d) calculation of many excited  states (including wavefunctions); 
(e) application to combined electron-nuclear motion quantum many-body problem.   These latter two require 
 $N_B \times N_B$ matrix \emph{diagonalization} (as opposed to just calculation of the ground state),
as well as the ability to go beyond  traditional SD representations.  Frontier (e) is, in addition, of special
relevance for QCC.

In this study, we have sought to demonstrate the potential advantages presented by the CCS approach 
in both classical and quantum computing contexts.  On the classical side, 
the CCS tensor-SOP decomposition of the wavefunction [\eq{tensorpsi}] enables astronomical basis sizes to be
considered with minimal RAM requirements---e.g., $N_B \approx 10^{15}$ for the harmonium calculation performed
in the present work, and up to $N_B \approx10^{18}$ more generally.  Of even greater import is the CCS tensor-SOP 
decomposition of the Hamiltonian operator itself, which reduces the complexity of both the classical 
and quantum calculations---essentially by restricting the set of coordinates that must be considered at one time, 
from six down to two.  

It is  worth discussing the quantum or CCS QCC case a bit further. In  Sec.~\ref{quantumCCS}, 
we have presented two different CCS Trotter implementations---i.e., the \emph{exact non-diagonal sinc} option,
and the  \emph{approximate diagonal sinc-DVR} option.   Both of these  represent an improvement over
the standard canonical Trotter QCC implementation, in that: (i) only two coordinates need be dealt with at a time
(potentially implying fewer quantum gates), and; (ii) quadrature error is reduced or eliminated entirely (implying
fewer basis functions or qubits).  In  comparing the two CCS QCC implementations to each other,  however,
we find that  the exact non-diagonal implementation  requires more quantum gates, but fewer 
qubits, than the diagonal sinc-DVR alternative---thereby  presenting a  useful ``engineering tradeoff.''
On the other hand, 
the computational bottleneck in either case is likely to be   evaluation of the Gaussian or exponential
function---which, like the inverse square-root, is also currently very expensive on quantum 
computers.\cite{babbushcomm}  Indeed, this cost could be so great as to render the various potential
advantages of the CCS approach (e.g. the reduction 
in circuit complexity that might otherwise occur) largely moot, in practice.
 
In a companion paper,\cite{exponential}  we will present a new algorithm for evaluating Gaussian
and exponential functions efficiently on quantum computers.  The cost is  the same as that
of performing a fairly small number of multiplications (with or without error 
correction),\cite{draper00,florio04,gidney19} and is thus 
on a par with the other costs associated with the present CCS QCC algorithm.   Running together
with the new exponentiation algorithm, then, the present approach 
would appear to become a highly competitive contender for first-quantized QCC.     

We conclude by once again stressing  the key point that \emph{the full
benefits of the CCS approach have yet to be realized.}  This is because to date, all calculations
have been performed using plane-wave and/or sinc basis functions, which represent a terrible
choice for Coulomb systems.  As discussed, something like $L\approx 100$ is required for such
systems, corresponding to $M=L^3=10^6$ single-particle orbitals.  For harmonium, plane waves
are a reasonably good---but still not great---choice of basis. Here, we find a reduction to $L=17$ or
$M\approx 5000$.  For the uniform or correlated electron gas application, plane waves are of 
course a much better choice, giving rise to $L=11$, or $M \approx 1000$ orbitals.  All of these
are requirements for achieving chemical accuracy. 

The main lesson here is the unsurprising one that \emph{a suitable choice of CCS basis for a given application
can lead to a dramatic reduction in basis size}. Ultimately, this should result in explicit classical calculations up 
to perhaps $N=6$.  Larger calculations could also be performed, using the output of a CCS calculation as a 
highly-efficient, multi-particle basis set (i.e., the $N$-particle, CCS version of a geminal basis set).  
Of course in practice, this will first require the development of  the ``optimal
CCS basis'' (OCCSB) functions of the \eq{CCSbasis} form---i.e., the CCS analog of HF molecular orbitals.  

As a technical matter, we advocate use of optimal separable 
basis theory\cite{poirier97b,poirier99qcII,poirier00prec}   to compute the OCCSB
functions (rather than traditional SCF), because the result is an optimized,  complete orthonormal  basis obtained
``all at once'' (as opposed to a collection of non-orthogonal ground and excited state SCF functions
obtained one at a time).  In any event, and as mentioned several times, the individual OCCSB functions
will incorporate correlation implicitly---thus potentially dramatically reducing the required basis size, even in 
cases where electron correlation is severe.  At the same time, the CCS tensor decomposition enables 
classical computers to handle $N_B = 10^{15}$ or more basis functions, which is very promising.

Of course, OCCSB calculations have yet to be performed on classical computers, the era of first-quantized 
QCC  is  still a ways off, and it is never certain what the future will bring.  Nevertheless, one thing
already seems quite clear: it is best to have a variety of different methods in one's computational arsenal---as 
indeed, has been the case within the confines of
 traditional second-quantized classical quantum chemistry  for decades.

\section*{Conflicts of interest}
There are no conflicts to declare.

\section*{Acknowledgements}
This work was  supported under contract from the US Army Research Office, Chemical Sciences Division
(W911NF1910023). The Robert A. Welch Foundation (D-1523) also provided support.  We also gratefully acknowledge 
a series of stimulating conversations with Ryan Babbush of Google Research. This technology is currently patent
pending by the Texas Tech University System. 



\balance




\providecommand*{\mcitethebibliography}{\thebibliography}
\csname @ifundefined\endcsname{endmcitethebibliography}
{\let\endmcitethebibliography\endthebibliography}{}


\end{document}